\documentclass[journal]{IEEEtranTCOM}
\normalsize

\usepackage[T1]{fontenc}
\ifCLASSINFOpdf
\else
\fi
\usepackage{amsmath,amssymb}
\usepackage{graphicx}
\usepackage{amsthm}
\theoremstyle{definition}

\newtheorem*{theorem*}{Theorem}

\usepackage{txfonts}
\usepackage{setspace}
\begin{document}
%
\title{Non-Linear Programming: Maximize SNR\\ for Designing Spreading Sequence -- Part II:\\Conditions for Optimal Spreading Sequences}

\author{Hirofumi~Tsuda ~\IEEEmembership{Student Member,~IEEE,} Ken Umeno
\thanks{H. Tsuda and K. Umeno are with the Department
of Applied Mathematics and Physics, Graduate School of Informatics, Kyoto University, Kyoto, 606-8561 Japan (email: tsuda.hirofumi.38u@st.kyoto-u.ac.jp, umeno.ken.8z@kyoto-u.ac.jp).}}


%

\markboth{IEEE Transactions on Communications}%
{Submitted paper}


\maketitle

\begin{abstract}
Signal to Noise Ratio (SNR) is an important index for wireless communications. In CDMA systems, spreading sequences are utilized. This series of papers show the method to derive spreading sequences as the solutions of non-linear programming: maximize SNR. In this paper, we derive the optimization problems with the expression SNR derived in Part I and the necessary conditions for the global solutions. We numerically solve the problems and evaluate the solutions with two factors, mean-square correlations and maximum mean-square correlations.
\end{abstract}


%
\IEEEpeerreviewmaketitle

\section{Introduction}
\IEEEPARstart{T}{here} are many works for designing spreading sequences to improve DS-CDMA systems \cite{dscdma}, which is one of the multiple access systems \cite{multiple}. In communication systems, it is important to increase their capacity \cite{shannon} to send information with high-rate. It is necessary and sufficient for achieving the spectral efficiency to increase the Signal to Noise Ratio (SNR) \cite{efficiency}. Bit Error Rate (BER) is also important index and it shows the quality of communication systems. It is known that BER becomes smaller when SNR becomes higher \cite{pursley}. Thus, it is important to increase SNR for communication systems.

The current spreading sequences are the Gold codes \cite{gold}. As the works for designing ones, we refer the reader to \cite{chaos_cdma}-\cite{indoor}. Their approaches to obtain spreading sequences are to design the system which generates sequences. However, it is not known what is necessary for systems to generate spreading sequences whose SNR is high. Therefore, it is not straightforward to design the desired systems. 

Our approach to design spreading sequences is to derive directly them as the solutions of optimizing problems: maximize SNR. Thus, our method has the advantage that our spreading sequences are guaranteed to have high SNR since they are the solutions of the problems. In particular, our optimization problems belong to non-linear programming problems \cite{nonlinear}. Non-linear programming has been developed and used for solving different problems. The necessary conditions for the global solution of a non-linear programming problem is known as the Karush-Kuhn-Tucker (KKT) condition \cite{KKT}. There are many methods to solve numerically non-linear programming problems, for example \cite{sqp}-\cite{slsqp}. 

Our optimization problems have parameters $Z_{i,k}$ $(i \neq k)$ and $Z_{i,i}$ which are corresponding to autocorrelation and crosscorrelation, that is, faded signal and interference noise, respectively. Thus, we can design flexibly spreading sequences with changing the parameters. The spreading sequences which are the solutions of the problems have resistance to fading effects when $Z_{i,i}$ becomes large and ones have resistance to interference noise when $Z_{i,k}$ $(i \neq k)$ becomes large. This result is shown as the numerical result.

In Part I \cite{part_I}, we have derived the new expression of SNR. With this expression, we construct two types of optimization problems: maximize the lower bound of the average of SNR and maximize minimum SNR in all the users. With our expression, we can numerically solve the problems and obtain the solutions. In general, there are relations between SNR and mean-square correlation \cite{fzcfam} \cite{meansquare}. Thus, we evaluate the solutions with two factors, mean-square correlations and maximum mean-square correlations and discuss the relation between our two types optimization problems and the factors. 

\section{Decomposition of Spreading Sequence and New Expression of SNR}
In Part I \cite{part_I}, we have shown how to decompose spreading sequences and derived the new expression of SNR. Let $\mathbf{s}_{k}$ be the spreading sequence of the user $k$ which is defined as
\begin{equation}
\mathbf{s}_k = (s_{k,1}, s_{k,2}, \ldots, s_{k,N})^\mathrm{T},
\end{equation}
where $N$ is the length of spreading sequences and $\mathbf{z}^\mathrm{T}$ is the transpose of $\mathbf{z}$. We decompose the spreading sequence $\mathbf{s}_{k}$ as
\begin{equation}
\mathbf{s}_k = \frac{1}{\sqrt{N}}\sum_{m=1}^N \alpha^{(k)}_m\mathbf{w}_m(0)=\frac{1}{\sqrt{N}}\sum_{m=1}^N \beta^{(k)}_m\mathbf{w}_m\left(\frac{1}{2N}\right),
\end{equation}
where $\mathbf{w}_m(\eta)$ is the basis vector whose $n$-th component is expressed as
\[\left(\mathbf{w}_m(\eta)\right)_n = \exp\left(2\pi j (n-1)\left(\frac{m}{N} + \eta\right)\right),\]
$j$ is an unit imaginary number, $\alpha_m^{(k)}$ and $\beta_m^{(k)}$ are the complex coefficients. Note that the each of vectors $\mathbf{w}_m(0)$ and $\mathbf{w}_m\left(\frac{1}{2N}\right)$ is orthogonal with respect to different $m$. Thus, the coefficients $\alpha^{(k)}_m$ and $\beta^{(k)}_m$ are obtained from $\mathbf{s}_{k}$. Then, the lower bound of SNR is written as
\begin{equation}
\operatorname{SNR}_i \geq \left\{\frac{1}{6N^2}\sum_{k=1}^K Z_{i,k} \sum_{m=1}^M S_m^{i,k} + \frac{N_0}{2PT}\right\}^{-1/2},
\label{eq:SNR}
\end{equation}
where
\begin{equation}
\begin{split}
S^{i,k}_m &= \left|\alpha_m^{(i)}\right|^2\left|\alpha_m^{(k)}\right|^2\left(1 + \frac{1}{2}\cos\left(2 \pi \frac{m}{N}\right)\right)\\
&+ \left|\beta_m^{(i)}\right|^2\left|\beta_m^{(k)}\right|^2\left(1 + \frac{1}{2}\cos\left(2 \pi \left(\frac{m}{N} + \frac{1}{2N}\right)\right)\right),
\end{split}
\label{eq:S}
\end{equation}
$Z_{i,k}$ $(i \neq k)$ and $Z_{i,i}$ are the coefficients of faded signal term and interference noise term, respectively, $K$ is the number of users, $P$ is the common power of the carrier signals, $T$ is the duration of one symbol and $N_0$ is the power of additive white Gaussian noise (AWGN). The equality is attained if the covariance function of the fading random process, $g_i(\tau)$ is the rectangular function, which is the worst case that the variance of the faded signal term is maximized.

\section{Optimization Problems for SNR}
In this section, our goal is to derive necessary conditions for the optimal spreading sequence which maximizes SNR for the case where the $g_i(\tau)$ is the worst. Here, the parameters $N$, $K$, $Z_{i,k}$ $(i \neq k)$ and $Z_{i,i}$ are fixed. We treat $Z_{i,k}$ as the weights among all the users. We ignore the Gaussian noise term since it has no relation to spreading sequences. To maximize the lower bound of SNR of the user $i$, we should minimize the first term of the denominator of Eq. (\ref{eq:SNR}). We consider the optimization problem $(\tilde{P})$
\begin{equation}
\begin{split}
(\tilde{P}) & \hspace{3mm} \min \hspace{2mm}\sum_{k=1}^K Z_{i,k} \sum_{m=1}^N S_m^{i,k}\\
\mbox{subject to}  \hspace{3mm} & {\boldsymbol \alpha^{(k)}} = \Phi {\boldsymbol \beta^{(k)}} \hspace{3mm}(k=1,2,\ldots,K),\\
& {\boldsymbol \beta^{(k)}} = \hat{\Phi} {\boldsymbol \alpha^{(k)}} \hspace{3mm}(k=1,2,\ldots,K),\\
&\left\|\boldsymbol \alpha^{(k)}\right\|^2 = \left\|\boldsymbol \beta^{(k)}\right\|^2 = N \hspace{3mm}(k=1,2,\ldots,K),
\end{split}
\end{equation}
where $\|\mathbf{x}\|$ is the Euclidian norm of the vector $\mathbf{x}$,
\begin{equation}
 {\boldsymbol \alpha^{(k)}} = \left( \begin{array}{c}
 \alpha_1^{(k)}\\
 \alpha_2^{(k)}\\
 \vdots\\
 \alpha_N^{(k)}
 \end{array} \right),  {\boldsymbol \beta^{(k)}} = \left( \begin{array}{c}
 \beta_1^{(k)}\\
 \beta_2^{(k)}\\
 \vdots\\
 \beta_N^{(k)}
 \end{array} \right),
  \end{equation}
  $\Phi$ and $\hat{\Phi}$ are the unitary matrices whose $(m,n)$-th components are
  \begin{equation}
  \begin{split}
   \Phi_{m,n}=&\frac{1}{N}\cdot\frac{2}{1-\exp(2 \pi j (\frac{n-m}{N} + \frac{1}{2N}))},\\
 \hat{\Phi}_{m,n}  =&\frac{1}{N}\cdot\frac{2}{1-\exp(2 \pi j (\frac{n-m}{N} - \frac{1}{2N}))}.
  \end{split}
  \end{equation}
The first two constraint terms have been discussed in \cite{basis}. The last constraint term is a signal power constraint. As can be seen in \cite{part_I}, this constraint is obtained from the condition that
\begin{equation}
\sum_{n=1}^N \left|s_{k,n}\right|^2 = N.
\end{equation}
In the problem $(\tilde{P})$, we take into account only the user $i$. However, we should consider all the users for designing spreading sequences. We show the two types of optimization problems: (i) maximize the lower bound of the average of SNR and (ii) maximize the minimum SNR in all the users.

\subsection{Maximize Average of SNR}
First, we consider the problem that we maximize the sum of $\operatorname{SNR}_i$. However, it is not obvious to obtain the global solution since the derivative of objective function is complicated. The denominator of SNR is important factor to increase SNR. Then, we consider the problem $(\tilde{P1}')$ which consists of the sum of the squared denominator of $\operatorname{SNR}_i$
\begin{equation}
\begin{split}
(\tilde{P1}')& \hspace{3mm} \min \hspace{2mm}\sum_{i=1}^K\sum_{k=1}^K Z_{i,k} \sum_{m=1}^N S_m^{i,k}\\
\mbox{subject to}  \hspace{3mm} & {\boldsymbol \alpha^{(k)}} = \Phi {\boldsymbol \beta^{(k)}} \hspace{3mm}(k=1,2,\ldots,K),\\
& {\boldsymbol \beta^{(k)}} = \hat{\Phi} {\boldsymbol \alpha^{(k)}} \hspace{3mm}(k=1,2,\ldots,K),\\
&\left\|\boldsymbol \alpha^{(k)}\right\|^2 = \left\|\boldsymbol \beta^{(k)}\right\|^2 = N \hspace{3mm}(k=1,2,\ldots,K).
\end{split}
\end{equation}
Note that the objective function has the bounds:
\begin{equation}
\begin{split}
&KN\left\{ Z_{AC,L}(R_{AC}+1) + Z_{CC,L}(K-1)R_{CC}\right\}\\
\leq &\sum_{i=1}^K\sum_{k=1}^K Z_{i,k} \sum_{m=1}^N S_m^{i,k}\\
\leq & 3KN\{ Z_{AC,U}(R_{AC}+1) + Z_{CC,U}(K-1)R_{CC}\},
\end{split}
\end{equation}
where 
\begin{equation}
\begin{split}
Z_{AC,L} &= \min_{i}Z_{i,i}, Z_{AC,U} = \max_{i}Z_{i,i},\\
Z_{CC,L} &= \min_{i \neq k}Z_{i,k}, Z_{CC,U} = \max_{i \neq k}Z_{i,k},
\end{split}
\end{equation}
and $R_{AC}$ and $R_{CC}$ are the mean-square autocorrelation and crosscorrelation, which are defined in Part I \cite{part_I}. There is the relation between the sum of the squared denominator of $\operatorname{SNR}_i$ and the sum of $\operatorname{SNR}_i$:
\begin{equation}
\begin{split}
\frac{1}{K}\sum_{i=1}^K \operatorname{SNR}_i  &\geq \frac{1}{\sqrt{K}}\left\{\frac{1}{K}\sum_{i=1}^K \left(\operatorname{SNR}_i \right)^2\right\}^{1/2}\\
&\geq \left[\frac{1}{\sum_{i=1}^K \{\operatorname{Denom}(\operatorname{SNR}_i )\}^2}\right]^{1/2},
\end{split}
\end{equation}
where $\operatorname{Denom}\{\operatorname{SNR}_i \}$ is the denominator of $\operatorname{SNR}_i$. This result is obtained from the relation between the arithmetic mean and the harmonic mean. In the problem $(\tilde{P1}')$, we evaluates the lower bound of the sum of $\operatorname{SNR}_i$.

In general, the variables of optimization problems are real numbers. However, the variables of the problem $(\tilde{P1}')$ is complex numbers. We rewrite the problem $(\tilde{P1}')$ to the real number optimization problem. To this end, we can use the method in \cite{cmgc} that transforms a complex-number vector to a real-number vector and a complex-number unitary matrix to a real-number orthogonal matrix. With these results, we have the problem $(P1')$
\begin{equation}
\begin{split}
(P1')& \hspace{3mm} \min \hspace{2mm}\sum_{i=1}^K\sum_{k=1}^K Z_{i,k} \sum_{m=1}^N \hat{S}_m^{i,k}\\
\mbox{subject to}  \hspace{3mm} & {\boldsymbol \alpha'^{(k)}} = \Phi' {\boldsymbol \beta'^{(k)}} \hspace{3mm}(k=1,2,\ldots,K),\\
& {\boldsymbol \beta'^{(k)}} = \hat{\Phi}' {\boldsymbol \alpha'^{(k)}} \hspace{3mm}(k=1,2,\ldots,K),\\
&\left\|\boldsymbol \alpha'^{(k)}\right\|^2 = \left\|\boldsymbol \beta'^{(k)}\right\|^2 = N \hspace{3mm}(k=1,2,\ldots,K),
\end{split}
\end{equation}
where
\begin{equation}
\begin{split}
\Phi' &= \left( \begin{array}{c c}
\operatorname{Re}[\Phi] & -\operatorname{Im}[\Phi]\\
\operatorname{Im}[\Phi] & \operatorname{Re}[\Phi]
\end{array} \right), \\
\hat{\Phi}' &= \left( \begin{array}{c c}
\operatorname{Re}[\hat{\Phi}] & -\operatorname{Im}[\hat{\Phi}]\\
\operatorname{Im}[\hat{\Phi}] & \operatorname{Re}[\hat{\Phi}]
\end{array} \right), \\
{\boldsymbol \alpha}'^{(k)} &= 
\left( \begin{array}{c}
{\boldsymbol \alpha}^{(k)} _{1}\\
{\boldsymbol \alpha}^{(k)} _{2}
\end{array} \right)
=\left( \begin{array}{c}
\operatorname{Re}[{\boldsymbol \alpha}^{(k)}]\\
\operatorname{Im}[{\boldsymbol \alpha}^{(k)}]
\end{array} \right),\\
 {\boldsymbol \beta}'^{(k)}  &= 
\left( \begin{array}{c}
{\boldsymbol \beta}^{(k)} _{1}\\
{\boldsymbol \beta}^{(k)} _{2}
\end{array} \right) = 
\left( \begin{array}{c}
\operatorname{Re}[{\boldsymbol \beta}^{(k)}]\\
\operatorname{Im}[{\boldsymbol \beta}^{(k)}]
\end{array} \right)
\end{split}
\end{equation}
and
\begin{equation}
\begin{split}
\hat{S}_m^{i,k}&= \left(\left(\alpha^{(i)}_{1,m}\right)^2 +\left(\alpha^{(i)}_{2,m}\right)^2  \right) \left(\left(\alpha^{(k)}_{1,m}\right)^2 +\left(\alpha^{(k)}_{2,m}\right)^2  \right)\\
&\cdot \left(1 + \frac{1}{2}\cos\left(2\pi\frac{m}{N}\right)\right)\\
&+\left(\left(\beta^{(i)}_{1,m}\right)^2 +\left(\beta^{(i)}_{2,m}\right)^2  \right)\left(\left(\beta^{(k)}_{1,m}\right)^2 +\left(\beta^{(k)}_{2,m}\right)^2  \right)\\
&\cdot \left(1 + \frac{1}{2}\cos\left(2\pi\left(\frac{m}{N} + \frac{1}{2N}\right)\right)\right).
 \end{split}
\end{equation}
The real values $\alpha^{(k)}_{1,m}$, $\alpha^{(k)}_{2,m}$, $\beta^{(k)}_{1,m}$ and $\beta^{(k)}_{2,m}$ are the $m$-th elements of ${\boldsymbol \alpha}^{(k)}_{1}$, ${\boldsymbol \alpha}^{(k)}_{2}$, ${\boldsymbol \beta}^{(k)}_{1}$ and ${\boldsymbol \beta}^{(k)}_{2}$.
We can reduce the two linear constraints to the one constraint, and the two norm constraints to the one norm constraint since $\hat{\Phi}'$ is an orthogonal matrix. We then obtain the problem $(P1)$
\begin{equation}
\begin{split}
(P1) &\hspace{3mm} \min \hspace{2mm}\sum_{i=1}^K\sum_{k=1}^K Z_{i,k} \sum_{m=1}^N \hat{S}_m^{i,k}\\
\mbox{subject to}  \hspace{3mm} & {\boldsymbol \beta'^{(k)}} = \hat{\Phi}' {\boldsymbol \alpha'^{(k)}} \hspace{3mm}(k=1,2,\ldots,K),\\
&\left\|\boldsymbol \alpha'^{(k)}\right\|^2 = N \hspace{3mm}(k=1,2,\ldots,K),
\end{split}
\end{equation}
We collect the variables of the problem $(P1)$ into $\mathbf{x}$. The vector $\mathbf{x}$ whose dimension is $4NK$ is expressed as
\begin{equation}
\mathbf{x} = \left( \begin{array}{c} 
\mathbf{x}_1\\
\mathbf{x}_2\\
\vdots\\
\mathbf{x}_K
\end{array} \right), \hspace{3mm}\mathbf{x}_k = \left( \begin{array}{c}
\boldsymbol\alpha'^{(k)}\\
\boldsymbol\beta'^{(k)}
\end{array} \right).
\end{equation}
The problem $(P1)$ is a non-linear programming. There are many numerical methods for non-linear programmings, for example \cite{sqp} \cite{nop}.\\

We show the necessary conditions for the global solution of problem $(P1)$. To this end, we focus on the KKT condition, which is known to be the necessary conditions for the global solutions \cite{KKT}. \\
To differentiate the objective function and the constraint functions of the problem $(P1)$, we express them as
\begin{equation}
\begin{split}
f(\mathbf{x}) &= \sum_{i=1}^K\sum_{k=1}^K Z_{i,k} \sum_{m=1}^N \hat{S}_m^{i,k},\\
c^{(k)}(\mathbf{x}) &=  {\boldsymbol \beta'^{(k)}} - \hat{\Phi}' {\boldsymbol \alpha'^{(k)}}\hspace{3mm} (k=1,2,\ldots,K),\\
d^{(k)}(\mathbf{x}) &= \left\|\boldsymbol \alpha'^{(k)}\right\|^2 - N\hspace{3mm} (k=1,2,\ldots,K).
\end{split}
\end{equation}
The constraint function $c^{(k)}(\mathbf{x})$ is a vector valued function. We divide the condition $c^{(k)}(\mathbf{x})$ into $2N$ conditions
\begin{equation}
\begin{split}
c^{(k)}_{1,m}(\mathbf{x}) &= \beta^{(k)}_{1,m} - \frac{1}{N}\left\{\sum_{n=1}^N \alpha^{(k)}_{1,n} - \sum_{n=1}^N \alpha^{(k)}_{2,n}\hat{\phi}_{m,n}\right\}, \\
c^{(k)}_{2,m}(\mathbf{x}) &= \beta^{(k)}_{2,m} - \frac{1}{N}\left\{ \sum_{n=1}^N \alpha^{(k)}_{1,n}\hat{\phi}_{m,n}  + \sum_{n=1}^N \alpha^{(k)}_{2,n} \right\},
\end{split}
\end{equation}
for $(m=1,2,\dots,N)$. In the above equations, $\hat{\phi}_{m,n}$ is 
\begin{equation}
\hat{\phi}_{m,n} = \frac{\sin\left(2 \pi \left(\frac{n-m}{N} - \frac{1}{2N}\right)\right)}{1- \cos\left(2 \pi \left(\frac{n-m}{N} - \frac{1}{2N}\right)\right)}
\end{equation}
and we have used the result that
\begin{equation}
\begin{split}
\operatorname{Re}[\hat{\Phi}_{m,n}] &= \frac{1}{N},\\
\operatorname{Im}[\hat{\Phi}_{m,n}] &= \frac{1}{N} \cdot \frac{\sin\left(2 \pi \left(\frac{n-m}{N} - \frac{1}{2N}\right)\right)}{1- \cos\left(2 \pi \left(\frac{n-m}{N} - \frac{1}{2N}\right)\right)}.\\
\end{split}
\end{equation}
We focus on the user $p$. For the user $p$, $c^{(k)}_{1,m}(\mathbf{x})$, $c^{(k)}_{2,m}(\mathbf{x})$ and $d^{(k)}(\mathbf{x})$ have no relation when $k \neq p$. For this reason, it is sufficient to consider only $c^{(p)}_{1,m}(\mathbf{x})$, $c^{(p)}_{2,m}(\mathbf{x})$ and $d^{(p)}(\mathbf{x})$ as the constraint functions. We define the integer $q$ $(1\leq q \leq N)$. Each of the derivatives of $f$ is
\begin{equation}
\begin{split}
\frac{\partial f(\mathbf{x})}{\partial \alpha^{(p)}_{1,q}} &= 2\alpha^{(p)}_{1,q}\left\{\sum_{i=1}^K Z_{i,p} A^{(i)}_q + \sum_{k=1}^K Z_{p,k} A^{(k)}_q\right\},\\
\frac{\partial f(\mathbf{x})}{\partial \alpha^{(p)}_{2,q}} &= 2\alpha^{(p)}_{2,q}\left\{\sum_{i=1}^K Z_{i,p} A^{(i)}_q + \sum_{k=1}^K Z_{p,k} A^{(k)}_q\right\},\\
\frac{\partial f(\mathbf{x})}{\partial \beta^{(p)}_{1,q}} &= 2\beta^{(p)}_{1,q}\left\{\sum_{i=1}^K Z_{i,p} B^{(i)}_q + \sum_{k=1}^K Z_{p,k} B^{(k)}_q\right\},\\
\frac{\partial f(\mathbf{x})}{\partial \beta^{(p)}_{2,q}} &= 2\beta^{(p)}_{2,q}\left\{\sum_{i=1}^K Z_{i,p} B^{(i)}_q + \sum_{k=1}^K Z_{p,k} B^{(k)}_q\right\},\\
\end{split}
\end{equation}
where
\begin{equation}
\begin{split}
A^{(k)}_q &= \left\{ \left(\alpha^{(k)}_{1,q}\right)^2 + \left(\alpha^{(k)}_{2,q}\right)^2 \right\}\left(1 + \frac{1}{2}\cos\left(2 \pi \frac{q}{N}\right)\right),\\
B^{(k)}_q &= \left\{ \left(\beta^{(k)}_{1,q}\right)^2 + \left(\beta^{(k)}_{2,q}\right)^2 \right\}\left(1 + \frac{1}{2}\cos\left(2 \pi \left(\frac{q}{N} + \frac{1}{2N}\right)\right)\right).
\end{split}
\end{equation}
Each of the derivative of $c^{(p)}_{1,m}(\mathbf{x})$, $c^{(p)}_{2,m}(\mathbf{x})$ and $d^{(p)}(\mathbf{x})$ is
\begin{equation}
\begin{split}
\frac{\partial c^{(p)}_{1,m}(\mathbf{x})}{\partial \alpha^{(p)}_{1,q}} &= -\frac{1}{N},\hspace{3mm}\frac{\partial c^{(p)}_{1,m}(\mathbf{x})}{\partial \alpha^{(p)}_{2,q}} = \frac{1}{N}\hat{\phi}_{m,q},\\
\frac{\partial c^{(p)}_{1,m}(\mathbf{x})}{\partial \beta^{(p)}_{1,q}} &= \delta_{mq},\hspace{3mm}\frac{\partial c^{(p)}_{1,m}(\mathbf{x})}{\partial \beta^{(p)}_{2,q}} = 0,\\
\end{split}
\label{eq:c_1}
\end{equation}
\begin{equation}
\begin{split}
\frac{\partial c^{(p)}_{2,m}(\mathbf{x})}{\partial \alpha^{(p)}_{1,q}} &=  -\frac{1}{N}\hat{\phi}_{m,q},\hspace{3mm}\frac{\partial c^{(p)}_{2,m}(\mathbf{x})}{\partial \alpha^{(p)}_{2,q}} =-\frac{1}{N},\\
\frac{\partial c^{(p)}_{2,m}(\mathbf{x})}{\partial \beta^{(p)}_{1,q}} &= 0,\hspace{3mm}\frac{\partial c^{(p)}_{2,m}(\mathbf{x})}{\partial \beta^{(p)}_{2,q}} = \delta_{mq},\\
\end{split}
\label{eq:c_2}
\end{equation}
\begin{equation}
\begin{split}
\frac{\partial d^{(p)}(\mathbf{x})}{\partial \alpha^{(p)}_{1,q}} &=  2\alpha^{(p)}_{1,q},\hspace{3mm}\frac{\partial d^{(p)}(\mathbf{x})}{\partial \alpha^{(p)}_{2,q}} =2\alpha^{(p)}_{2,q},\\
\frac{\partial d^{(p)}(\mathbf{x})}{\partial \beta^{(p)}_{1,q}} &= \frac{\partial d^{(p)}(\mathbf{x})}{\partial \beta^{(p)}_{2,q}} = 0.\\
\end{split}
\label{eq:d}
\end{equation}
We define $\lambda^{(k)}_{1,m}$, $\lambda^{(k)}_{2,m}$ and $\mu^{(k)}$ as the Lagrange multipliers for $c^{(k)}_{1,m}(\mathbf{x})$, $c^{(k)}_{2,m}(\mathbf{x})$ and $d^{(k)}(\mathbf{x})$. These variables satisfy
\begin{equation}
\begin{split}
&\nabla f(\mathbf{\tilde{x}}) + \sum_{k=1}^K \sum_{m=1}^N \left\{ \lambda^{(k)}_{1,m} \nabla c^{(k)}_{1,m}(\mathbf{\tilde{x}}) +  \lambda^{(k)}_{2,m}\nabla c^{(k)}_{2,m}(\mathbf{\tilde{x}})\right\}\\
&+   \sum_{k=1}^K \mu^{(k)} \nabla d^{(k)}(\mathbf{\tilde{x}}) = \mathbf{0},
\label{eq:KKT}
\end{split}
\end{equation}
where $\mathbf{\tilde{x}}$ is the global solution of the problem $(P1)$. Equation (\ref{eq:KKT}) is the KKT condition for the problem $(P1)$ and is the necessary condition that $\mathbf{\tilde{x}}$ is the global solution. If it is the case, then $\lambda^{(k)}_{1,m}$, $\lambda^{(k)}_{2,m}$ and $\mu^{(k)}$ are real numbers.

We define the each of the elements of the global solution $\mathbf{\tilde{x}}$ as $\tilde{\alpha}^{(k)}_{1,m}$, $\tilde{\alpha}^{(k)}_{2,m}$, $\tilde{\beta}^{(k)}_{1,m}$ and $\tilde{\beta}^{(k)}_{1,m}$. From $\mathbf{\tilde{x}}$, we can calculate the variable $\tilde{A}^{(k)}_q$ and $\tilde{B}^{(k)}_q$ which is defined as
\begin{equation}
\begin{split}
\tilde{A}^{(k)}_q &= \left\{ \left(\tilde{\alpha}^{(k)}_{1,q}\right)^2 + \left(\tilde{\alpha}^{(k)}_{2,q}\right)^2 \right\}\left(1 + \frac{1}{2}\cos\left(2 \pi \frac{q}{N}\right)\right),\\
\tilde{B}^{(k)}_q &= \left\{ \left(\tilde{\beta}^{(k)}_{1,q}\right)^2 + \left(\tilde{\beta}^{(k)}_{2,q}\right)^2 \right\}\left(1 + \frac{1}{2}\cos\left(2 \pi \left(\frac{q}{N} + \frac{1}{2N}\right)\right)\right).
\end{split}
\end{equation}

We focus on $\tilde{\beta}^{(p)}_{1,q}$ and $\tilde{\beta}^{(p)}_{2,q}$. From Eq. (\ref{eq:KKT}), $\tilde{\beta}^{(p)}_{1,q}$ and $\tilde{\beta}^{(p)}_{2,q}$ must satisfy
\begin{equation}
\begin{split}
\lambda^{(p)}_{1,q} =& -2\tilde{\beta}^{(p)}_{1,q}\left\{\sum_{i=1}^K Z_{i,p} \tilde{B}^{(i)}_q + \sum_{k=1}^K Z_{p,k} \tilde{B}^{(k)}_q\right\} ,\\
\lambda^{(p)}_{2,q} =& -2\tilde{\beta}^{(p)}_{2,q}\left\{\sum_{i=1}^K Z_{i,p} \tilde{B}^{(i)}_q + \sum_{k=1}^K Z_{p,k} \tilde{B}^{(k)}_q\right\}
\end{split}
\end{equation}
From the above equations, $\lambda^{(p)}_{1,q}$ and $\lambda^{(p)}_{2,q}$ are determined since $\tilde{\beta}^{(p)}_{1,q}$ and $\tilde{\beta}^{(p)}_{2,q}$ are given. Similarly, $\tilde{\alpha}^{(p)}_{1,q}$ and $\alpha^{(p)}_{2,q}$ must satisfy 
\begin{equation}
\begin{split}
&2N\tilde{\alpha}^{(p)}_{1,q}\left[ \mu^{(p)} + \left\{\sum_{i=1}^K Z_{i,p}\tilde{A}_q^{(i)} + \sum_{k=1}^K Z_{p,k}\tilde{A}_q^{(k)} \right\}\right]\\
 =& \sum_{m=1}^N \lambda_{1,m}^{(p)} + \sum_{m=1}^N \lambda_{2,m}^{(p)} \hat{\phi}_{m,q},\\
&2N\tilde{\alpha}^{(p)}_{2,q}\left[ \mu^{(p)} + \left\{\sum_{i=1}^K Z_{i,p}\tilde{A}_q^{(i)} + \sum_{k=1}^K Z_{p,k}\tilde{A}_q^{(k)} \right\}\right]\\
 =& -\sum_{m=1}^N \lambda_{1,m}^{(p)} \hat{\phi}_{m,q} + \sum_{m=1}^N \lambda_{2,m}^{(p)}
\end{split}
\label{eq:necessary}
\end{equation}
for all $q$. The Lagrange multiplier $\mu^{(p)}$ is the common variable in the above $2N$ equations. It is the necessary condition for the global solution $\mathbf{\tilde{x}}$ whether $\mu^{(p)}$ exists which satisfies Eq. (\ref{eq:necessary}).\\

\subsection{Maximize Minimum SNR}
Finally, instead of considering the optimization problem: maximize the lower bound of the average of SNR, we consider the problem: maximize the minimum SNR in all the users. This is expressed as the problem $(\tilde{P2}')$
\begin{equation}
\begin{split}
(\tilde{P2}')& \hspace{3mm} \min \max_{i} \sum_{k=1}^K Z_{i,k} \sum_{m=1}^N S_m^{i,k}\\
\mbox{subject to}  \hspace{3mm} & {\boldsymbol \alpha^{(k)}} = \Phi {\boldsymbol \beta^{(k)}} \hspace{3mm}(k=1,2,\ldots,K),\\
& {\boldsymbol \beta^{(k)}} = \hat{\Phi} {\boldsymbol \alpha^{(k)}} \hspace{3mm}(k=1,2,\ldots,K),\\
&\left\|\boldsymbol \alpha^{(k)}\right\|^2 = \left\|\boldsymbol \beta^{(k)}\right\|^2 = N \hspace{3mm}(k=1,2,\ldots,K).
\end{split}
\end{equation}
This problem is a mini-max problem \cite{minimax}. Note that the objective function has the bound that
\begin{equation}
\begin{split}
&\max_{i} \sum_{k=1}^K Z_{i,k} \sum_{m=1}^N S_m^{i,k}\\
\leq & 3N\left\{Z_{AC,U}\left(\max_{i}R^{(i)}_{AC} + 1\right) + Z_{CC,U}\left(K-1\right)\left(\max_{i}R^{(i)}_{CC}\right)\right\},
\end{split}
\end{equation}
where $R^{(i)}_{AC}$ and $R^{(i)}_{CC}$ are the mean-square autocorrelation and crosscorrelation of the user $i$, which is defined in \cite{part_I}. We call $\max_{i} R^{(i)}_{AC}$ and $\max_{i} R^{(i)}_{CC}$ as maximum mean-square autocorrelation and crosscorrelation, respectively. 

The general method to solve the mini-max problem is to introduce a slack variable $t$. The problem $(\tilde{P2}')$ is rewritten as the problem $(P2')$,
\begin{equation}
\begin{split}
(P2')& \hspace{3mm} \min t \\
\mbox{subject to}  \hspace{3mm} & {\boldsymbol \alpha^{(k)}} = \Phi {\boldsymbol \beta^{(k)}} \hspace{3mm}(k=1,2,\ldots,K),\\
& {\boldsymbol \beta^{(k)}} = \hat{\Phi} {\boldsymbol \alpha^{(k)}} \hspace{3mm}(k=1,2,\ldots,K),\\
&\left\|\boldsymbol \alpha^{(k)}\right\|^2 = \left\|\boldsymbol \beta^{(k)}\right\|^2 = N \hspace{3mm}(k=1,2,\ldots,K),\\
&t \geq \sum_{k=1}^K Z_{i,k} \sum_{m=1}^N S_m^{i,k} \hspace{3mm}(i=1,2,\ldots,K).
\end{split}
\end{equation}
Similar to the problem $(P1)$ and $(P1')$, we rewrite $(P2')$ to the real-variable problem $(P2)$,
\begin{equation}
\begin{split}
(P2) &\hspace{3mm} \min t\\
\mbox{subject to}  \hspace{3mm} & {\boldsymbol \beta'^{(k)}} = \hat{\Phi}' {\boldsymbol \alpha'^{(k)}} \hspace{3mm}(k=1,2,\ldots,K),\\
&\left\|\boldsymbol \alpha'^{(k)}\right\|^2 = N \hspace{3mm}(k=1,2,\ldots,K),\\
&t \geq \sum_{k=1}^K Z_{i,k} \sum_{m=1}^N \hat{S}_m^{i,k} \hspace{3mm}(i=1,2,\ldots,K)
\end{split}
\end{equation}
We collect the variables of the problem $(P2)$ into $\mathbf{x}$. The vector $\mathbf{x}$ is expressed as
\begin{equation}
\mathbf{x} = \left( \begin{array}{c} 
t\\
\mathbf{x}_1\\
\mathbf{x}_2\\
\vdots\\
\mathbf{x}_K
\end{array} \right), \hspace{3mm}\mathbf{x}_k = \left( \begin{array}{c}
\boldsymbol\alpha'^{(k)}\\
\boldsymbol\beta'^{(k)}
\end{array} \right).
\end{equation}
The dimension of $\mathbf{x}$ is $4NK+1$.\\

We show the KKT conditions of the problem $(P2)$.
To differentiate the objective function and the constraint functions of the problem $(P2)$, we express them as
\begin{equation}
\begin{split}
f(\mathbf{x}) &= t,\\
c^{(k)}(\mathbf{x}) &=  {\boldsymbol \beta'^{(k)}} - \hat{\Phi}' {\boldsymbol \alpha'^{(k)}}\hspace{3mm} (k=1,2,\ldots,K),\\
d^{(k)}(\mathbf{x}) &= \left\|\boldsymbol \alpha'^{(k)}\right\|^2 - N\hspace{3mm} (k=1,2,\ldots,K),\\
e^{(i)}(\mathbf{x})  &=  \sum_{k=1}^K Z_{i,k} \sum_{m=1}^N \hat{S}_m^{i,k} - t\hspace{3mm} (i=1,2,\ldots,K).
\end{split}
\label{eq:cons2}
\end{equation}
Similar to the problem $(P1)$, we divide the condition $c^{(k)}(\mathbf{x})$ into $2N$ conditions $c^{(k)}_{1,m}(\mathbf{x})$ and $c^{(k)}_{2,m}(\mathbf{x})$. It is clear that $\nabla f$ is a vector function that
\begin{equation}
\nabla f = \left( \begin{array}{c c c c}
1 & 0 & \cdots & 0
\end{array} \right)^{\mathrm{T}}.
\end{equation}
The vector function $\nabla f$ has $1$ for the first element and $0$ for the other elements. The vector functions $\nabla c^{(k)}_{1,m}$, $\nabla c^{(k)}_{2,m}$ and $\nabla d^{(k)}$ are the same to the problem $(P1)$ except the derivative with respect to $t$. Note that
\begin{equation}
\frac{\partial c^{(k)}_{1,m}(\mathbf{x})}{\partial t} = \frac{\partial c^{(k)}_{2,m}(\mathbf{x})}{\partial t} = \frac{\partial d^{(k)}(\mathbf{x})}{\partial t} = 0.
\end{equation}
We consider the constraints associated with the user $p$. Each of the derivatives of $e^{(i)}$ is
 \begin{equation}
\begin{split}
\frac{\partial e^{(i)}(\mathbf{x})}{\partial t} &= -1,\\
\frac{\partial e^{(i)}(\mathbf{x})}{\partial \alpha^{(p)}_{1,q}} &= 2(1 + \delta_{ip})\alpha^{(p)}_{1,q}  Z_{i,p} A^{(i)}_q,\\
\frac{\partial e^{(i)}(\mathbf{x})}{\partial \alpha^{(p)}_{2,q}} &= 2(1 + \delta_{ip})\alpha^{(p)}_{2,q} Z_{i,p} A^{(i)}_q,\\
\frac{\partial e^{(i)}(\mathbf{x})}{\partial \beta^{(p)}_{1,q}} &= 2(1 + \delta_{ip})\beta^{(p)}_{1,q} Z_{i,p} B^{(i)}_q,\\
\frac{\partial e^{(i)}(\mathbf{x})}{\partial \beta^{(p)}_{2,q}} &= 2(1 + \delta_{ip})\beta^{(p)}_{2,q} Z_{i,p} B^{(i)}_q.\\
\end{split}
\end{equation}
 We define $\lambda^{(k)}_{1,m}$, $\lambda^{(k)}_{2,m}$, $\mu^{(k)}$ and $\nu^{(i)}$ as the Lagrange multipliers corresponding to $c^{(k)}_{1,m}(\mathbf{x})$, $c^{(k)}_{2,m}(\mathbf{x})$, $d^{(k)}(\mathbf{x})$ and $e^{(i)}(\mathbf{x})$. The KKT conditions of the problem $(P2)$ is expressed as
\begin{equation}
\begin{split}
&\nabla f(\mathbf{\tilde{x}}) + \sum_{k=1}^K \sum_{m=1}^N \left\{ \lambda^{(k)}_{1,m} \nabla c^{(k)}_{1,m}(\mathbf{\tilde{x}}) +  \lambda^{(k)}_{2,m}\nabla c^{(k)}_{2,m}(\mathbf{\tilde{x}})\right\}\\
&+   \sum_{k=1}^K \mu^{(k)} \nabla d^{(k)}(\mathbf{\tilde{x}}) + \sum_{i=1}^K \nu^{(i)} \nabla e^{(i)}(\mathbf{\tilde{x}})= \mathbf{0},
\label{eq:KKT2}
\end{split}
\end{equation}
where $\mathbf{\tilde{x}}$ is the global solution of the problem $(P2)$. The Lagrange multiplier $\nu^{(i)}$ has to satisfy
\begin{equation}
\nu^{(i)} \geq 0, \hspace{2mm}\mbox{and}\hspace{2mm} \nu^{(i)} = 0 \hspace{2mm}\mbox{for}\hspace{2mm}  e^{(i)}(\mathbf{\tilde{x}}) <0.
\label{eq:nu}
 \end{equation}
 The other Lagrange multipliers are real numbers.
 
Similar to $(P1)$, We define the each of the elements of the global solution $\mathbf{\tilde{x}}$ as $\tilde{t}$, $\tilde{\alpha}^{(k)}_{1,m}$, $\tilde{\alpha}^{(k)}_{2,m}$, $\tilde{\beta}^{(k)}_{1,m}$, $\tilde{\beta}^{(k)}_{1,m}$, and the variables as $\tilde{A}^{(k)}_q$ and  $\tilde{B}^{(k)}_q$. It is clear that the global solution $\tilde{t}$ corresponds the squared denominator of minimum SNR. Therefore, all the users are divided into two sets, one is the set of users whose SNR equals $\tilde{t}^{-1/2}$ and the other is the set of users whose SNR is higher than $\tilde{t}^{-1/2}$. Then, we define the set of the users who have the minimum SNR
 \begin{equation}
 U = \left\{i \in \{1,2,\ldots,K\} \middle| \sum_{k=1}^K Z_{i,k} \sum_{m=1}^N \hat{S}_m^{i,k} = \tilde{t}\right\}.
 \label{eq:U}
\end{equation}
From Eq. (\ref{eq:nu}) and Eq. (\ref{eq:U}), we obtain the relation
\begin{equation}
\nu^{(i)} = 0 \hspace{2mm}\mbox{for}\hspace{2mm} i \not\in U.
\label{eq:relation_U}
\end{equation}
Focusing on the variable $\tilde{t}$, we obtain the equation from Eqs. (\ref{eq:KKT2})-(\ref{eq:relation_U})
\begin{equation}
\sum_{i \in U} \nu^{(i)} = 1.
  \end{equation}
We consider the two cases, one is that $p \in U$ and the other is that $p \not\in U$.

First, we consider the case where the user $p \not\in U$. In this case, $\nu^{(p)}=0$. We focus on the variables $\tilde{\beta}^{(p)}_{1,m}$ and $\tilde{\beta}^{(p)}_{2,m}$ for ($m = 1,2,\ldots, N$). From Eqs. (\ref{eq:c_1}) (\ref{eq:c_2}) (\ref{eq:cons2}) and (\ref{eq:KKT2}), we obtain 
 \begin{equation}
  \lambda^{(p)}_{1,m} = \lambda^{(p)}_{2,m} = 0.
  \label{eq:condition1}
  \end{equation}
  From the above result, the constraint $c^{(p)}(\mathbf{x})$ has no relation to Eq. (\ref{eq:KKT2}). From Eq. (\ref{eq:d}) and Eq. (\ref{eq:KKT2}), considering the constraints $d^{(p)}(\mathbf{\tilde{x}})$, we obtain 
 \begin{equation}
  2\mu^{(p)} \tilde{\alpha}^{(p)}_{1,m} = 2\mu^{(p)} \tilde{\alpha}^{(p)}_{2,m} = 0.
  \end{equation}  
  The vector $\tilde{\boldsymbol \alpha}'^{(p)}$ is not the zero vector since it has to satisfy $\left\|\tilde{\boldsymbol \alpha}'^{(k)}\right\|^2 = N$, where $\tilde{\boldsymbol \alpha}'^{(k)}$ is the vector $\boldsymbol \alpha^{(k)}$ of the global solution. Then, we obtain the result
   \begin{equation}
  \mu^{(p)} = 0.
  \end{equation}
  From the above results, the user $p \not\in U$ has no relation to Eq. (\ref{eq:KKT2}).
  
 Finally, we consider the case where the user $p \in U$. From Eq. (\ref{eq:KKT2}), we obtain the four types of the equations
\begin{equation} 
\begin{split}
&-\frac{1}{N}\sum_{m=1}^N \left( \lambda^{(p)}_{1,m} +\lambda^{(p)}_{2,m}\hat{\phi}_{m,q}\right) + 2\mu^{(p)}\tilde{\alpha}^{(p)}_{1,q}\\
+& 2\nu^{(p)}\tilde{\alpha}^{(p)}_{1,q}\sum_{i=1}^K (1 + \delta_{ip})Z_{i,p} \tilde{A}^{(i)}_q=0,
     \end{split}
       \label{eq:condition2}
  \end{equation}
      \begin{equation} 
\begin{split}
&\frac{1}{N}\sum_{m=1}^N \left( \lambda^{(p)}_{1,m} \hat{\phi}_{m,q}-\lambda^{(p)}_{2,m}\right) + 2\mu^{(p)}\tilde{\alpha}^{(p)}_{2,q}\\
+& 2\nu^{(p)} \tilde{\alpha}^{(p)}_{2,q} \sum_{i=1}^K (1 + \delta_{ip})Z_{i,p}\tilde{A}^{(i)}_q=0,
      \end{split}
        \label{eq:condition3}
  \end{equation}
    \begin{equation} 
\lambda^{(p)}_{1,q} + 2\nu^{(p)} \tilde{\beta}^{(p)}_{1,q} \sum_{i=1}^K (1 + \delta_{ip})Z_{i,p}\tilde{B}^{(i)}_q=0,
  \label{eq:condition4}
  \end{equation}
      \begin{equation} 
\lambda^{(p)}_{2,q} + 2\nu^{(p)} \tilde{\beta}^{(p)}_{2,q} \sum_{i=1}^K (1 + \delta_{ip})Z_{i,p}\tilde{B}^{(i)}_q=0
  \label{eq:condition5}
  \end{equation}
  for $q = 1,2,\ldots,N$ and $p \in U$. The KKT conditions of the problem $(P2)$ are Eqs. (\ref{eq:condition2})-(\ref{eq:condition5}).
  
\section{Numerical Result}
In this section, we set $N=31$ and $K=4$. We numerically obtain the solutions of the problem $(P1)$ and the problem $(P2)$ and compare them with other sequences, the Gold codes \cite{gold}, The FZC sequences \cite{zadoff} \cite{chu}, Sarwate's sequences \cite{sarwate}. Sarwate's sequences are expressed as
\begin{equation}
s_{k,n} = \exp\left(2 \pi j n \left( \frac{\sigma_k}{N}\right) \right),
\end{equation}
where $\sigma_k \in \{0,1,2,\ldots,N-1\}$ is an initial element assigned to the user $k$. In \cite{sarwate}, FZC sequences and Sarwate's sequences are proposed as the sequences which are special in Sarwate's limitation. The second peak of periodic autocorrelation of FZC sequences is zero and the periodic crosscorrelation of Sarwate's sequences is always zero. These sequences have the remarkable feature of correlation.

We treat all the users equally, that is, we choose the parameter $Z_{i,k}$ as
\begin{equation}
Z_{i,k} = \left\{ \begin{array}{c c}
Z_{AC} & i = k\\
Z_{CC} & i \neq k
\end{array} \right. ,
\end{equation}
where $Z_{AC}$ and $Z_{CC}$ are the positive numbers. We solve the problems $(P1)$ and $(P2)$ in the pairs of $(Z_{AC},Z_{CC}) = (1,2), (2,1)$ and compare the solutions. Through these comparisons, our another goal is to obtain the relation among our problems and two factors, mean-square correlations $R_{AC}$ and $R_{CC}$, and maximum mean-square correlation in users $\max_i R^{(i)}_{AC}$ and $\max_i R^{(i)}_{CC}$, which are defined in Part I. 

In the problem $(P1)$, we evaluate the value
\begin{equation}
\left\{\frac{1}{6N^2K}\sum_{i=1}^K\sum_{k=1}^K Z_{i,k} \sum_{m=1}^N \hat{S}_m^{i,k}\right\}^{-1/2}.
\label{eq:ave_SNR}
\end{equation}
On the other hand, in the problem $(P2)$, we evaluate the value
\begin{equation}
\left[\frac{1}{6N^2}\left\{ \max_{i} \sum_{k=1}^K Z_{i,k} \sum_{m=1}^N \hat{S}_m^{i,k}\right\}\right]^{-1/2}.
\label{eq:min_SNR}
\end{equation}
In Table \ref{table1}, we note the language and the libraries which we used. To obtain solutions, we used the algorithm, Sequential Least Squares Programming (SLSQP) \cite{slsqp}. We define the three types of errors 
\begin{equation}
\begin{split}
e_1 &= \max_{k} \max\left\{\left|N - \left\|\tilde{\boldsymbol \alpha}'^{(k)}\right\|^2 \right|, \left|N - \left\|\tilde{\boldsymbol \beta}'^{(k)}\right\|^2 \right| \right\},\\
e_2 &= \max_{k} \left\| \tilde{\boldsymbol \beta}'^{(k)} - \hat{\Phi}'\tilde{\boldsymbol \alpha}'^{(k)} \right\|_{\infty},\\
e_3 &=  \min_{i} \left\{t - \sum_{k=1}^K Z_{i,k} \sum_{m=1}^N \hat{S}_m^{i,k}\right\},
\end{split}
\label{eq:error}
\end{equation} 
where $\| \cdot \|_\infty$ is the sup norm. In particularly, the error $e_3$ is related to only the problem $(P2)$. These errors show whether the solutions satisfy the constraints of the problems. In Tables \ref{table2} and \ref{table3}, we note the maximum values of $e_1$ and $e_2$ and the minimum value of $e_3$ of each problem in our trials. These results show that our solutions satisfy the constraints.

\begin{table}[htbp]
\centering
\vspace{5mm}
\caption{Language and Library}
\label{table1}
\begin{tabular}{|c | c|}\hline
Language/Library & Version \\\hline\hline
Python & 3.4.3\\\hline
SciPy & 0.15.1\\\hline
NumPy & 1.9.2 \\\hline
\end{tabular}
\end{table}

\begin{table}[htbp]
\centering
\vspace{5mm}
\caption{Errors of Solutions: Problem $(P1)$}
\label{table2}
\begin{tabular}{|c || c | c|  c| c|}\hline
$(Z_{AC},Z_{CC})$& $(1,2)$ & $(2,1)$  \\\hline\hline
$e_1$& $9.23\times10^{-14}$ & $9.95 \times 10^{-14}$ \\\hline
$e_2$& 0.0 & 0.0 \\\hline
\end{tabular}
\end{table}

\begin{table}[htbp]
\centering
\vspace{5mm}
\caption{Errors of Solutions: Problem $(P2)$}
\label{table3}
\begin{tabular}{|c || c | c|  c| c|}\hline
$(Z_{AC},Z_{CC})$ & $(1,2)$ & $(2,1)$  \\\hline\hline
$e_1$& $8.53\times10^{-14} $& $9.59 \times 10^{-14}$ \\\hline
$e_2$& 0.0& 0.0 \\\hline
$e_3$& 0.0&0.0\\\hline
\end{tabular}
\end{table}

In the appendices A and B, numerical solutions of the problems $(P1)$ and $(P2)$ are shown. We plot 100000 solutions of each problem and the orange point as the solution whose SNR is the highest in our trials. Our solution do not exceed all the other sequences in SNR. In the both of the problems, the solutions which have higher SNR have lower mean square cross-correlation and auto-correlation. We obtain the similar solutions to Sarwate's sequences in terms of mean-square correlations $R_{AC}$ and $R_{CC}$, and maximum mean-square correlations $\max_i R^{(i)}_{AC}$ and $\max_i R^{(i)}_{CC}$. However, we do not obtain the similar to the Gold codes and the FZC sequences in terms of them. This result shows that the Gold codes and the FZC sequences may be singular solutions of the problems $(P1)$ and $(P2)$.

In Part I, we have discussed the relation between SNR and mean-square correlations. Moreover, in Section III, we discuss the relation between the problem $(P1)$ and the mean-square correlations, and one between the problem $(P2)$ and the maximum mean-square correlations. The positions of bluer points change when the ratio between $Z_{AC}$ and $Z_{CC}$ are varied. With $(Z_{AC},Z_{CC}) = (1,2)$, SNR of the solutions becomes higher when the mean-square crosscorrelation becomes lower. However, $(Z_{AC},Z_{CC}) = (2,1)$, SNR becomes higher when the mean-square autocorrelation becomes lower. These results show that the spreading sequences which are the solutions of the problems have resistance to fading effects when $Z_{i,i}$ becomes large. Similarly, the spreading sequences have resistance to interference noise when $Z_{i,k}$ $(i \neq k)$ becomes large. These trends can be seen in both of the factors, mean-square correlations and maximum mean-square correlations.

\begin{figure}[htbp] 
   \centering
   \includegraphics[width=2.3in]{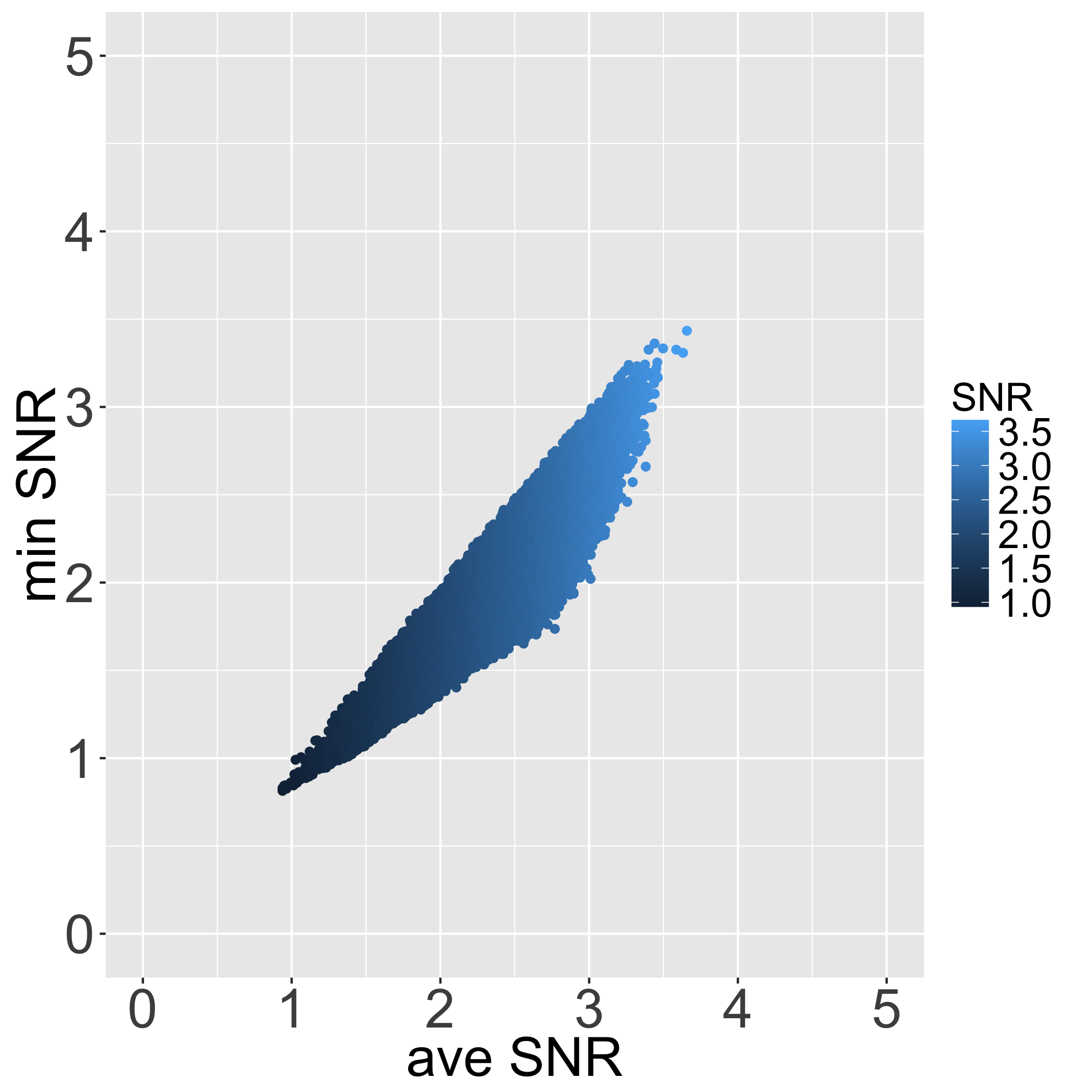} 
   \caption{Relation between average SNR and minimum SNR: $K=4, Z_{AC}=1, Z_{CC}=2$}
   \label{fig:4_1_2_relationP1}
\end{figure}

\begin{figure}[htbp] 
   \centering
   \includegraphics[width=2.3in]{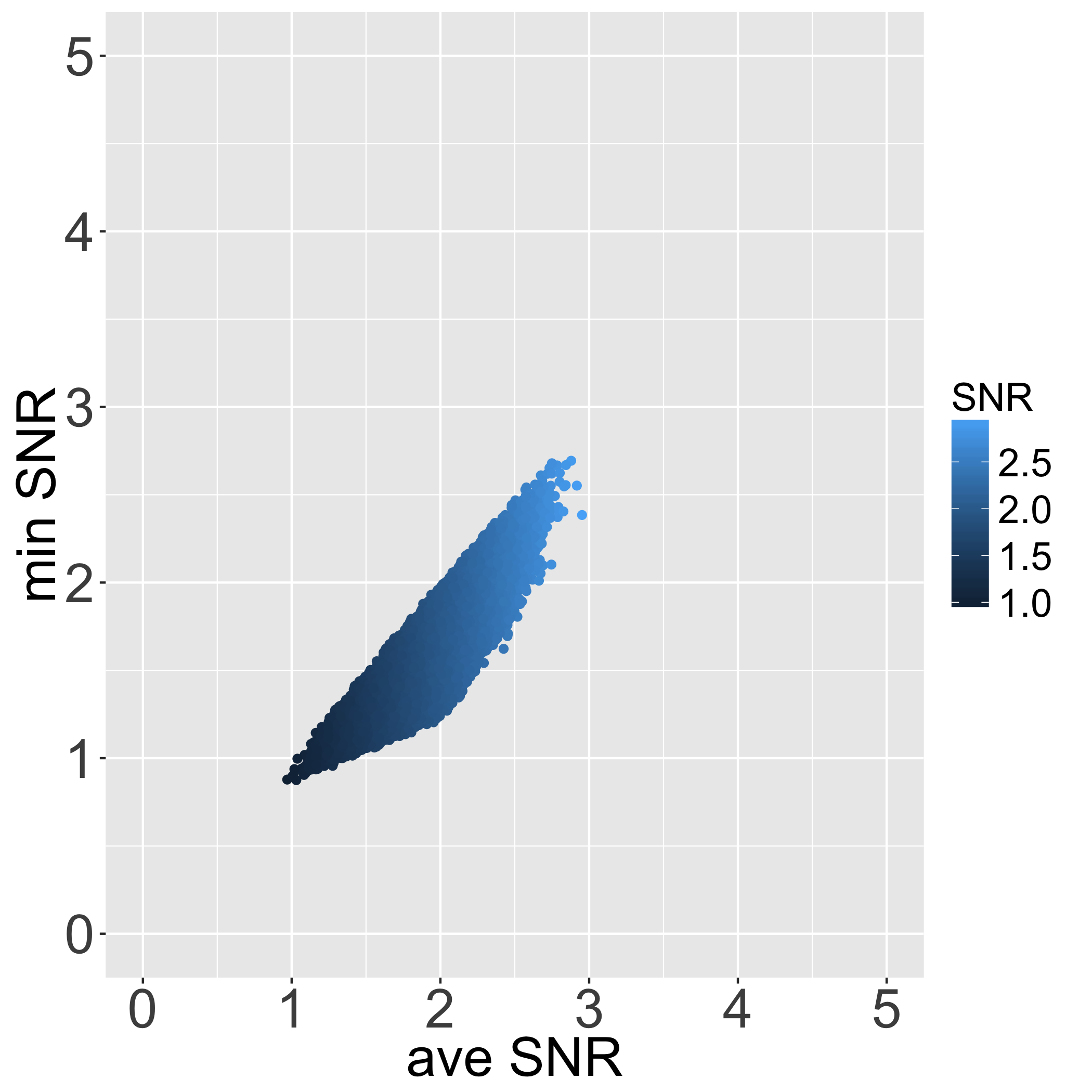} 
   \caption{Relation between average SNR and minimum SNR: $K=4, Z_{AC}=2, Z_{CC}=1$}
   \label{fig:4_2_1_relationP1}
\end{figure}

\begin{figure}[htbp] 
   \centering
   \includegraphics[width=2.3in]{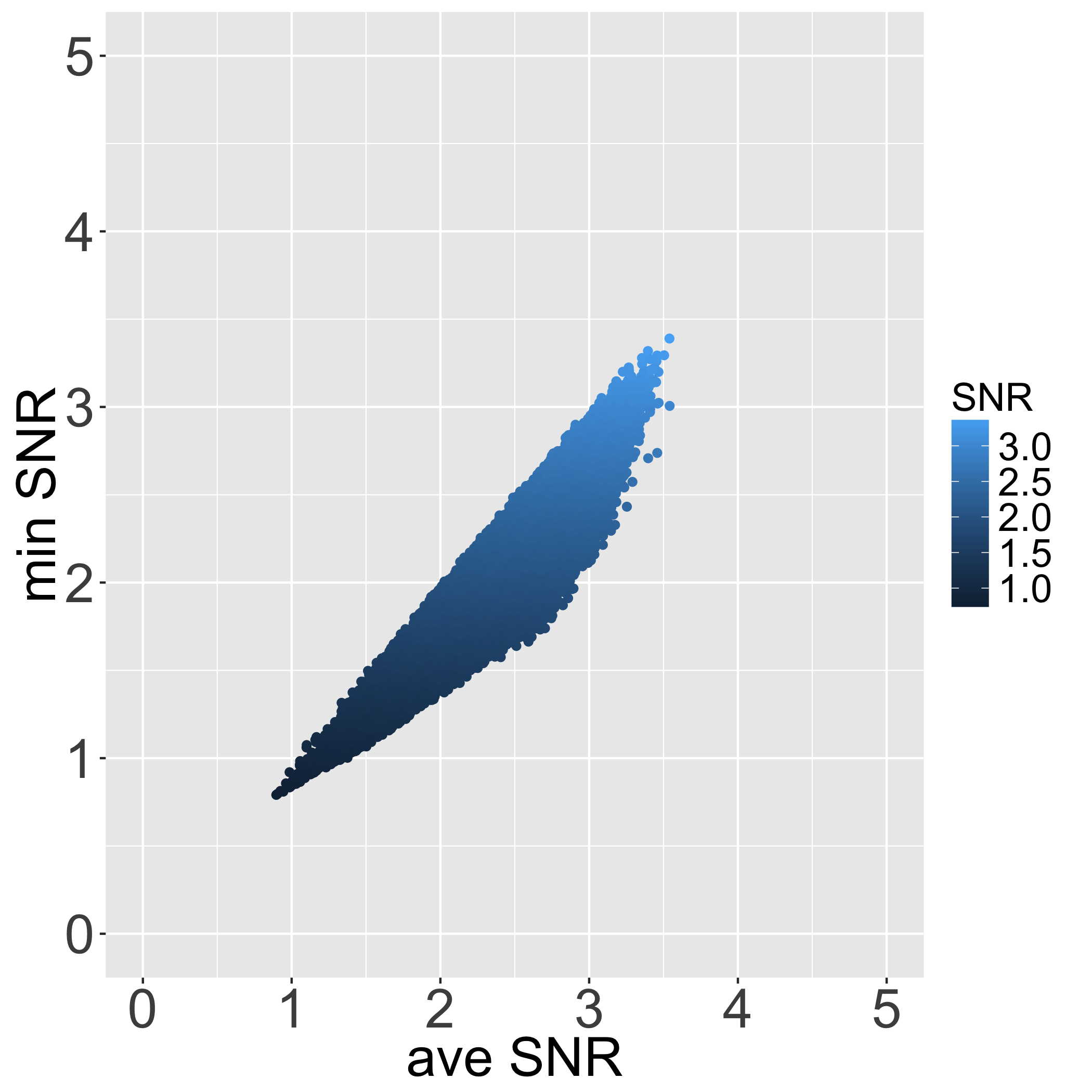} 
   \caption{Relation between average SNR and minimum SNR: $K=4, Z_{AC}=1, Z_{CC}=2$}
   \label{fig:4_1_2_relationP2}
\end{figure}

\begin{figure}[htbp] 
   \centering
   \includegraphics[width=2.3in]{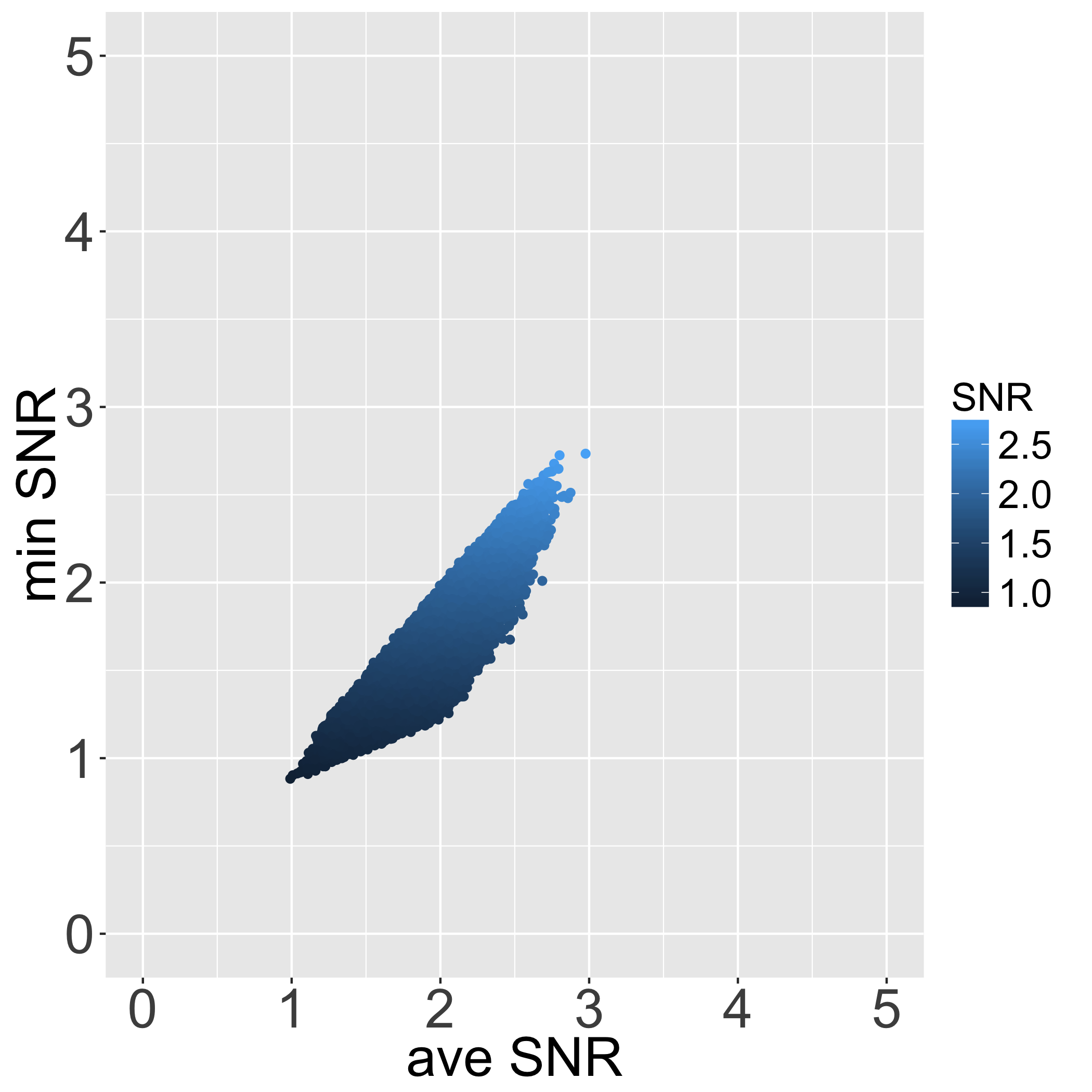} 
   \caption{Relation between average SNR and minimum SNR: $K=4, Z_{AC}=2, Z_{CC}=1$}
   \label{fig:4_2_1_relationP2}
\end{figure}

Figures \ref{fig:4_1_2_relationP1}, \ref{fig:4_2_1_relationP1}, \ref{fig:4_1_2_relationP2}, and \ref{fig:4_2_1_relationP2} show the relation between the average SNR (Eq. (\ref{eq:ave_SNR})) and the minimum SNR (Eq. (\ref{eq:min_SNR})) with the solutions of each problem. They show that the average SNR becomes close to the minimum SNR when SNR becomes high. This result implies that the global solutions of $(P1)$ may correspond to these of $(P2)$, that is, in both problems, SNR of each user may be the same.

\section{Conclusion}
We have shown the new expression of SNR formula. With this expression, we have constructed the two types of optimization problems: maximize the lower bound of SNR and maximize the minimum SNR. Then, we have derived necessary conditions of each problem. However, the global solutions of the problem $(P1)$ and $(P2)$ are not obtained since these problems are not convex programming. The better solutions will be found if the suitable algorithm is chosen, for example, the barrier function method \cite{barrier} and the penalty method \cite{penalty}. 

A remaining issue is to obtain the global solutions of the problems. These solutions guaranteed an optimality. The CDMA system will be developed if the global solutions are found.

\appendices

\section{Figures of Numerical Result in Problem $(P1)$}

\begin{figure}[htbp] 
   \centering
   \includegraphics[width=2.3in]{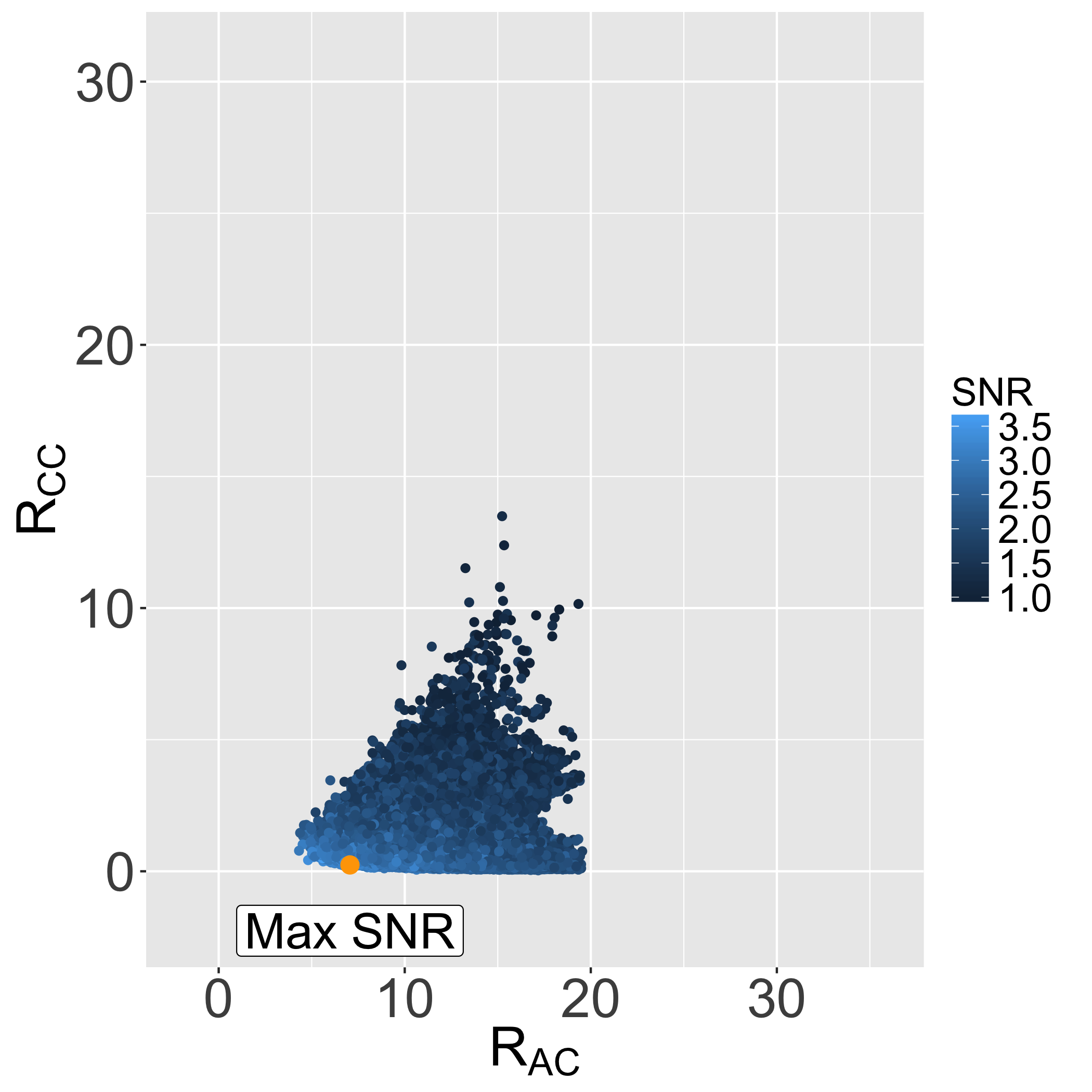} 
   \caption{Evaluation of solutions: mean-square correlation: $K=4, Z_{AC}=1, Z_{CC}=2$}
   \label{fig:4_1_2_r}
\end{figure}

\begin{figure}[htbp] 
   \centering
   \includegraphics[width=2.3in]{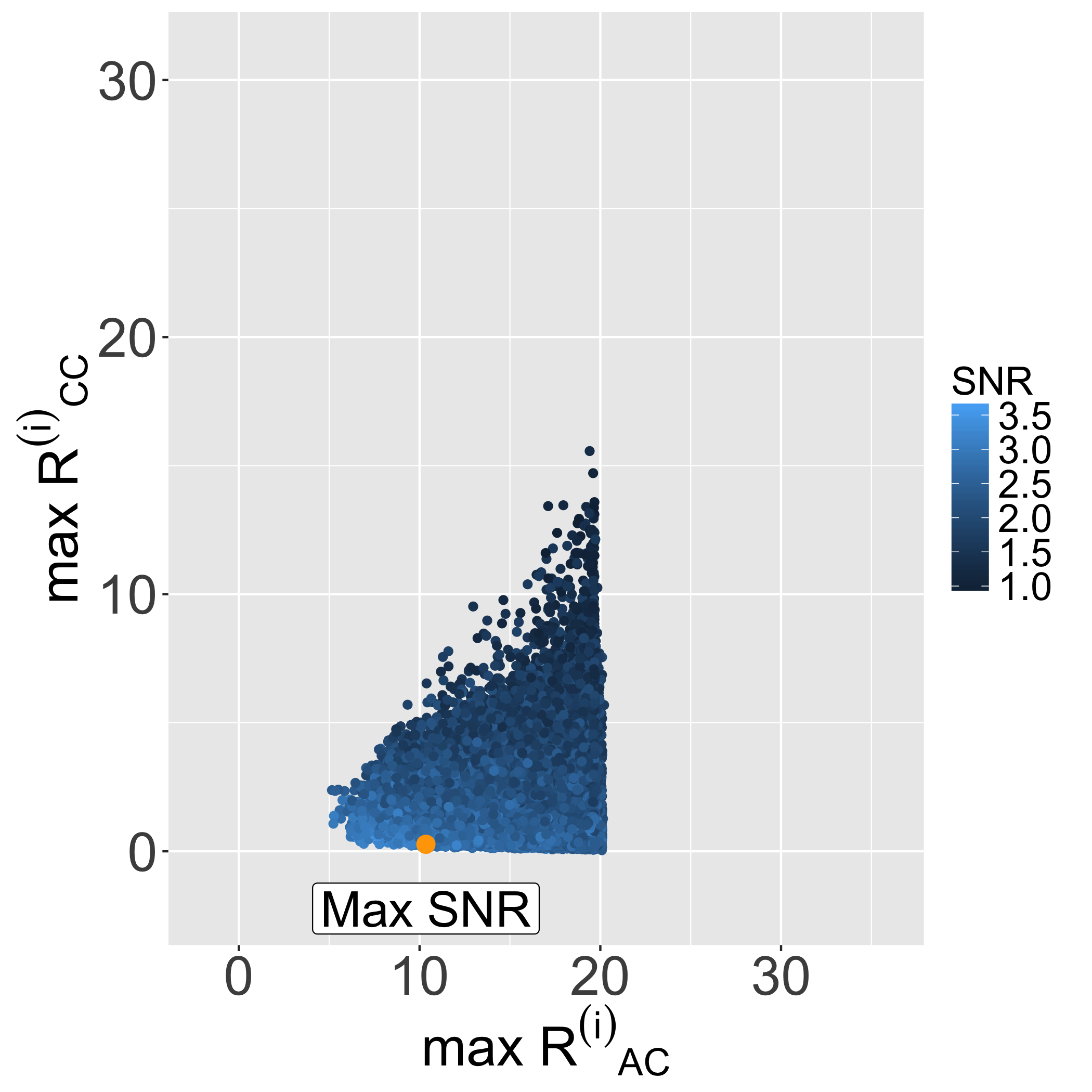} 
   \caption{Evaluation of solutions: maximum mean-square correlation in all the users: $K=4, Z_{AC}=1, Z_{CC}=2$}
   \label{fig:4_1_2_rmax}
\end{figure}

\begin{figure}[htbp] 
   \centering
   \includegraphics[width=2.3in]{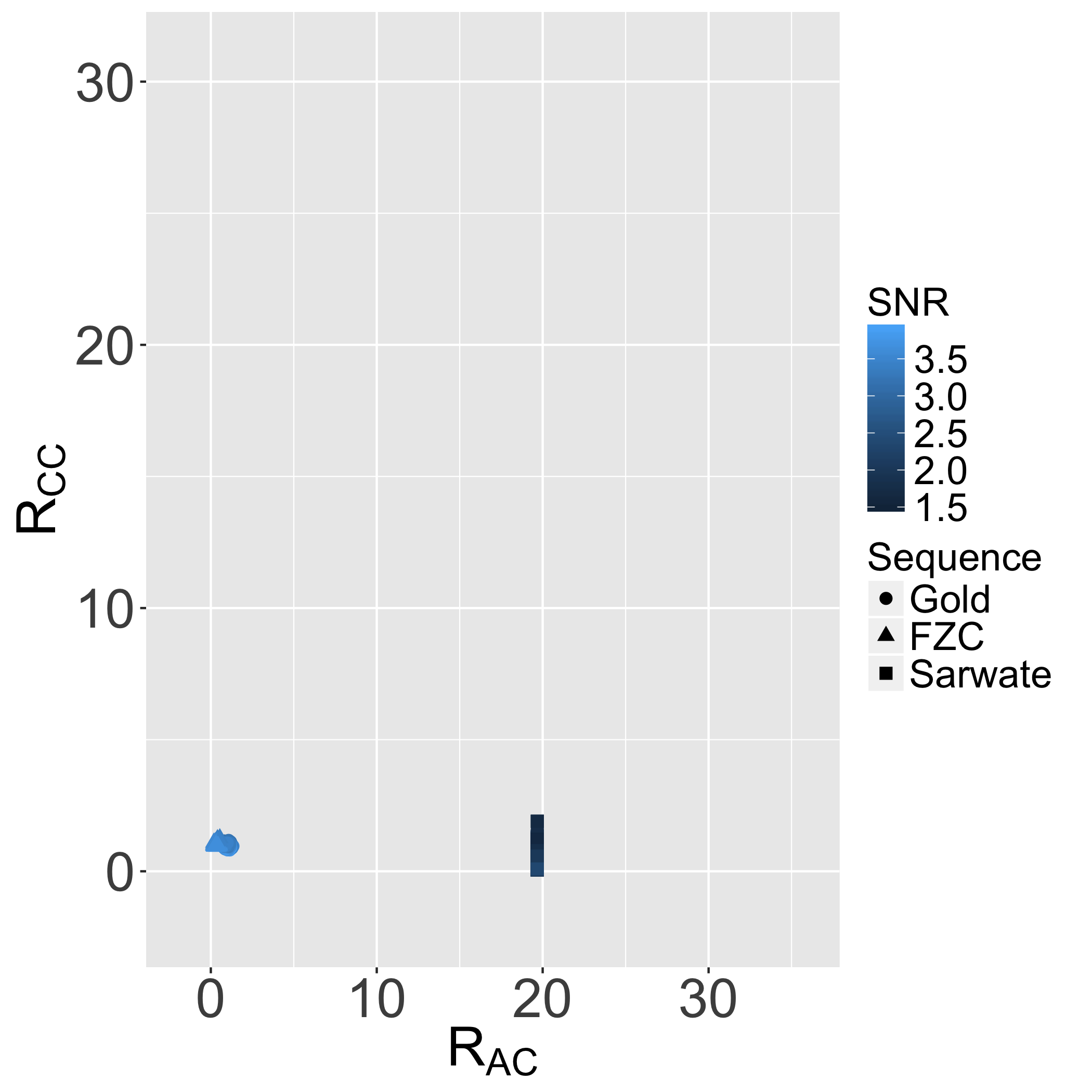} 
   \caption{Other sequences: mean-square correlation: $K=4, Z_{AC}=1, Z_{CC}=2$}
   \label{fig:4_1_2_other_r}
\end{figure}

\begin{figure}[htbp] 
   \centering
   \includegraphics[width=2.3in]{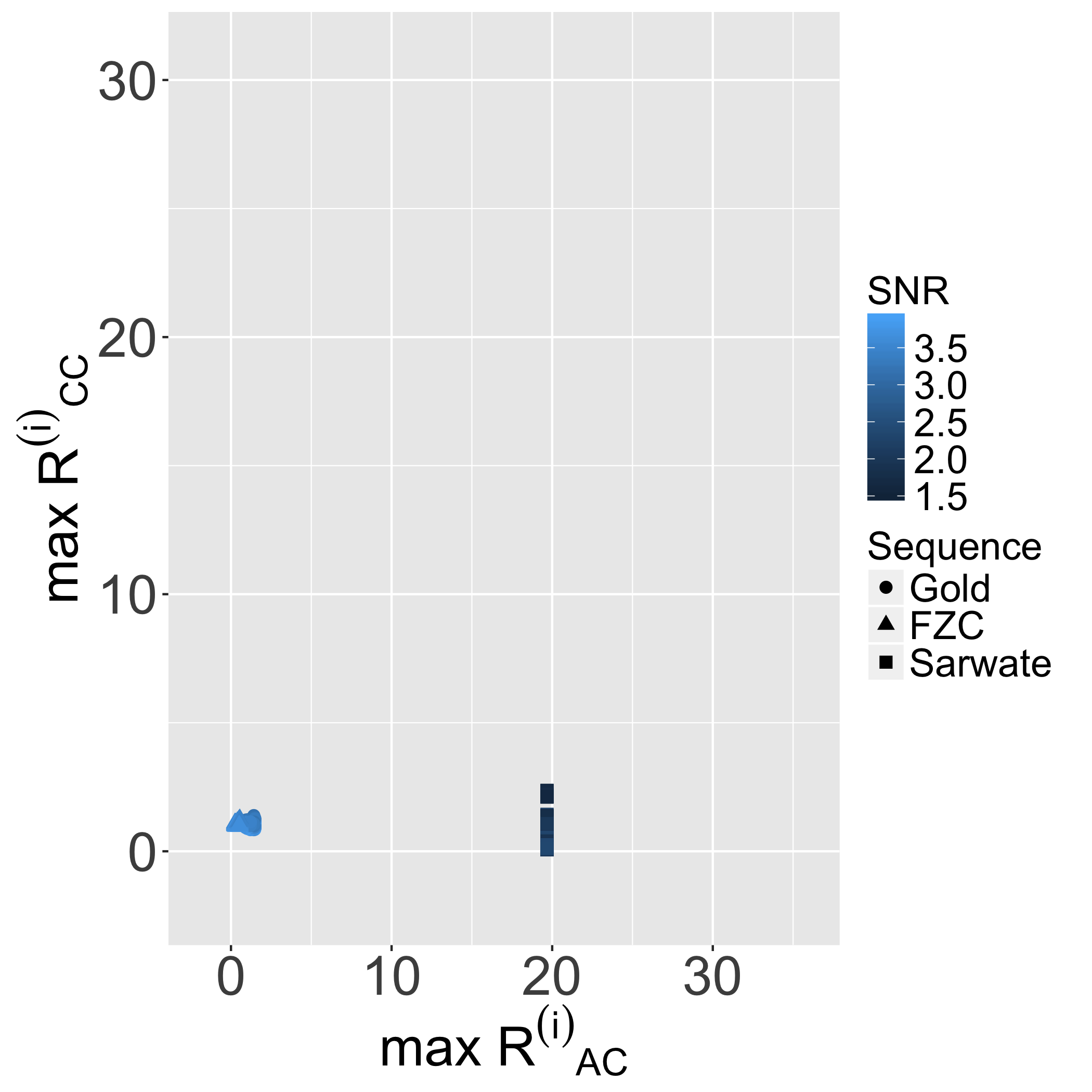} 
   \caption{Other sequences: maximum mean-square correlation in all the users: $K=4, Z_{AC}=1, Z_{CC}=2$}
   \label{fig:4_1_2_other_rmax}
\end{figure}

\begin{figure}[htbp] 
   \centering
   \includegraphics[width=2.3in]{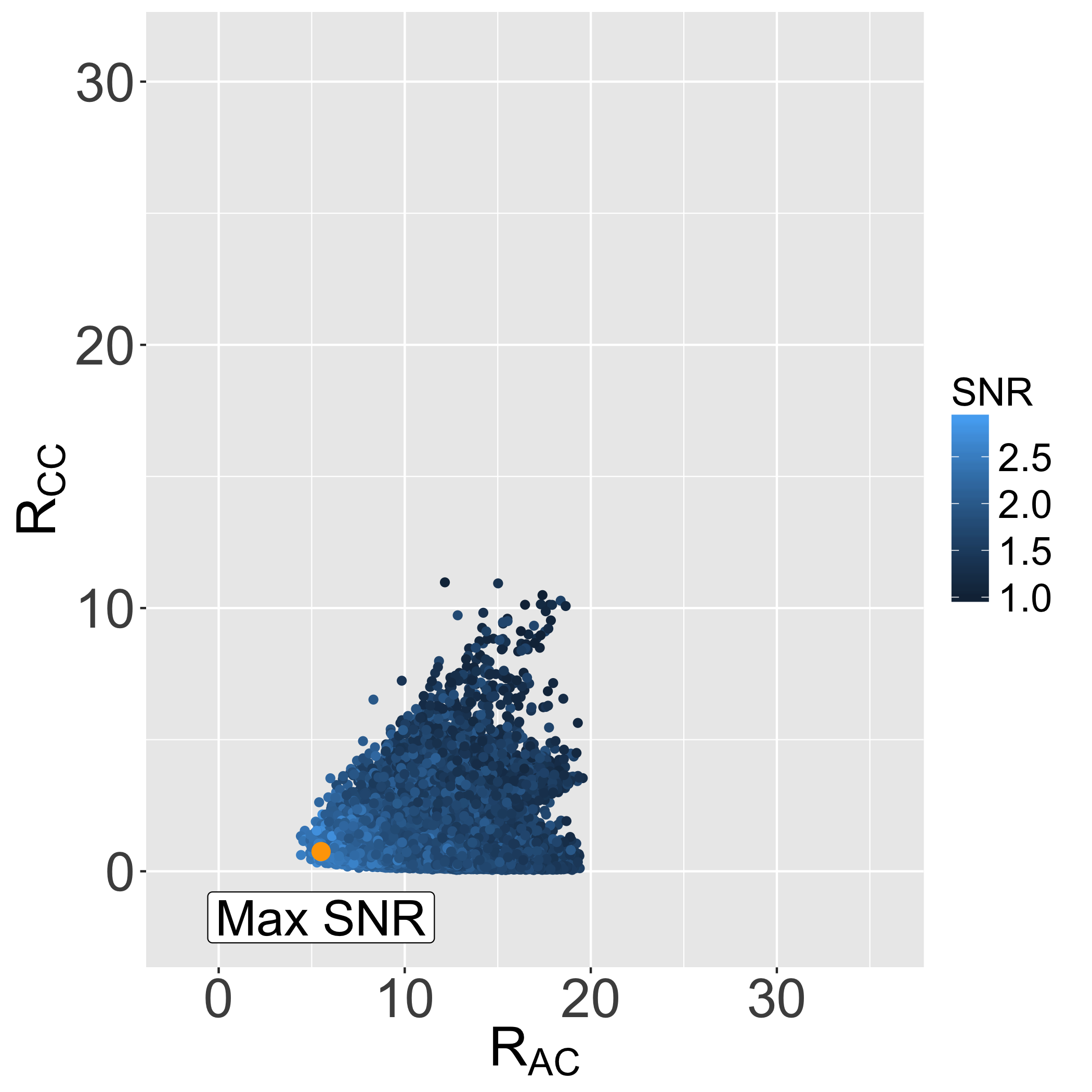} 
   \caption{Evaluation of solutions: mean-square correlation: $K=4, Z_{AC}=2, Z_{CC}=1$}
   \label{fig:4_2_1_r}
\end{figure}

\begin{figure}[htbp] 
   \centering
   \includegraphics[width=2.3in]{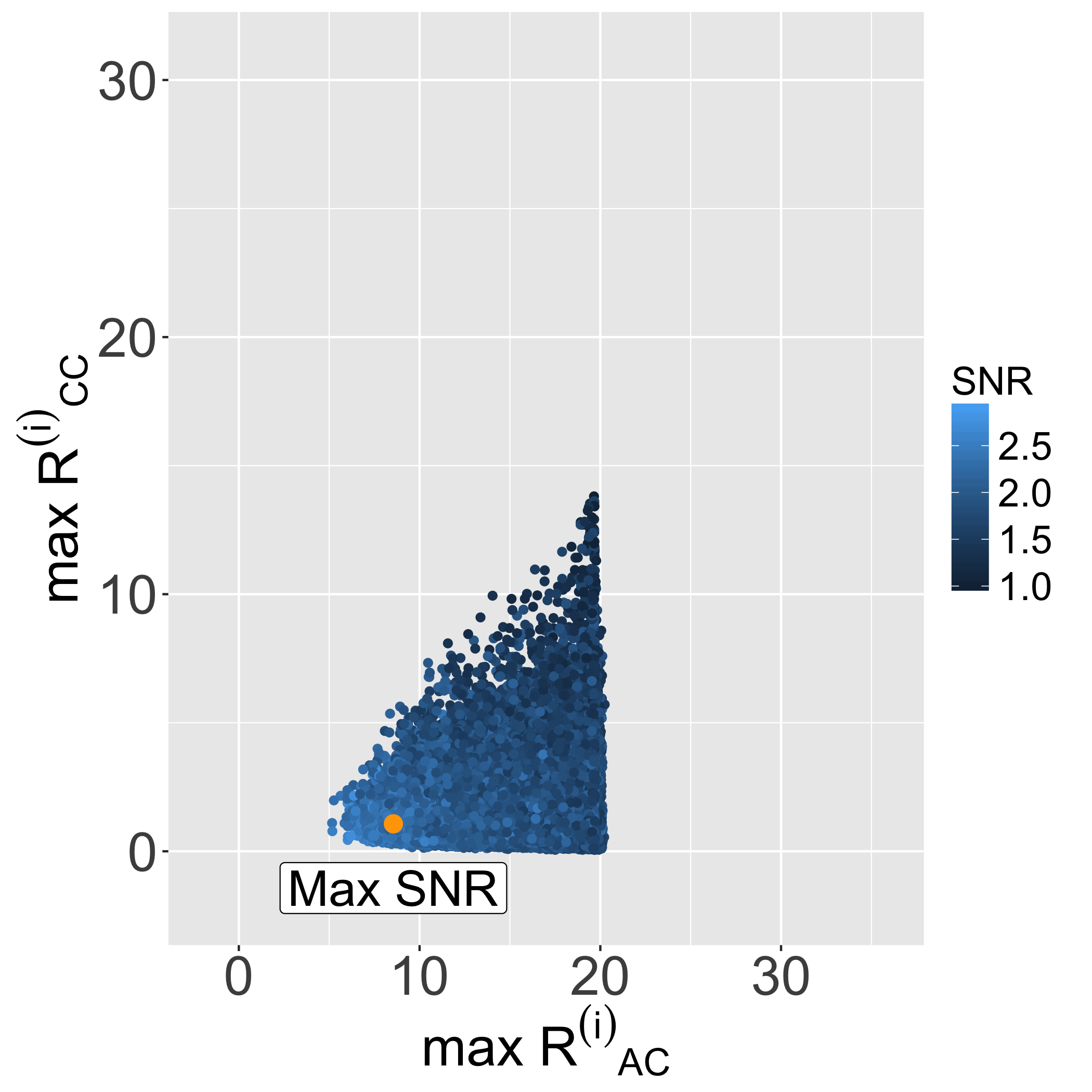} 
   \caption{Evaluation of solutions: maximum mean-square correlation in all the users: $K=4, Z_{AC}=2, Z_{CC}=1$}
   \label{fig:4_2_1_rmax}
\end{figure}

\begin{figure}[htbp] 
   \centering
   \includegraphics[width=2.3in]{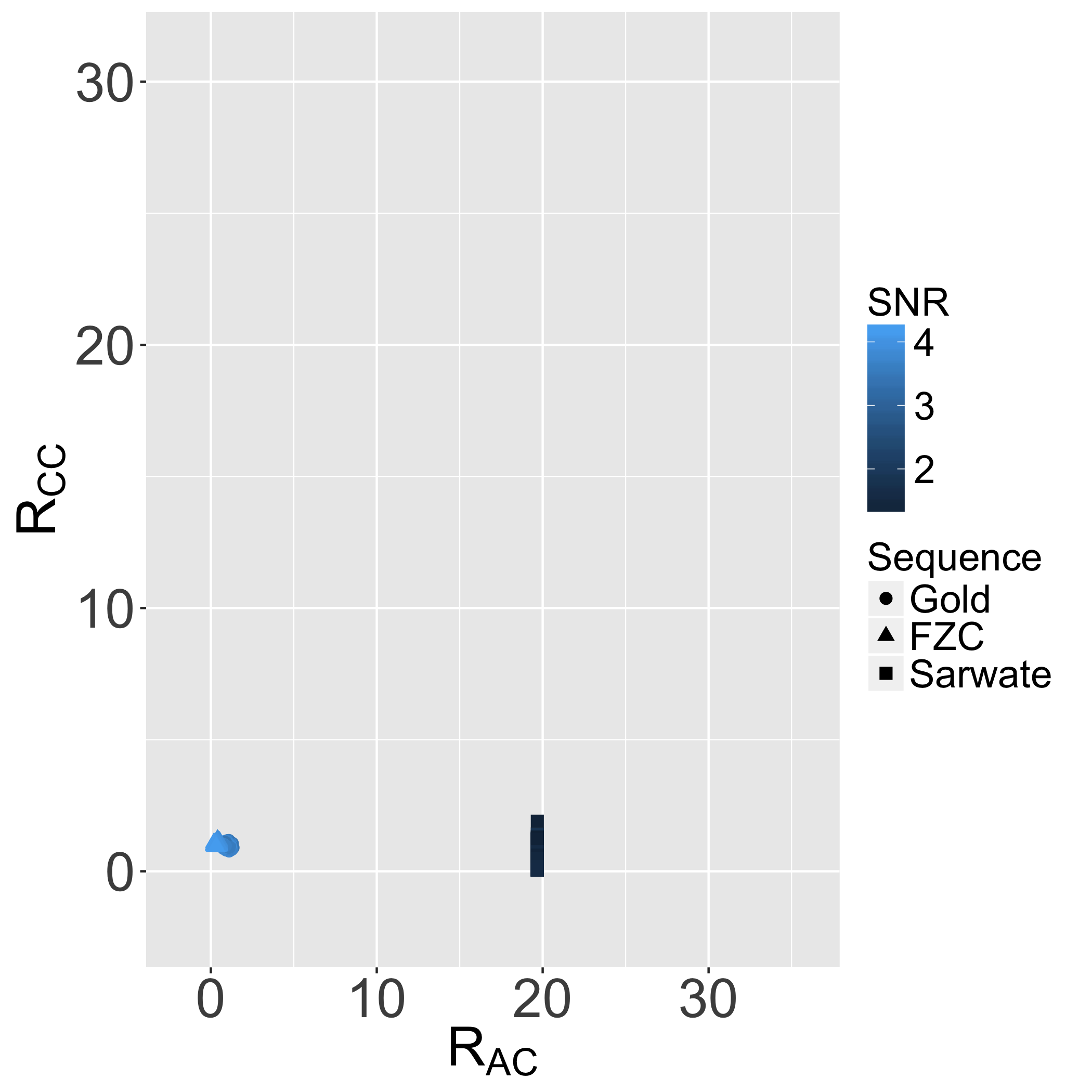} 
   \caption{Other sequences: mean-square correlation: $K=4, Z_{AC}=2, Z_{CC}=1$}
   \label{fig:4_1_1_other_r}
\end{figure}

\begin{figure}[htbp] 
   \centering
   \includegraphics[width=2.3in]{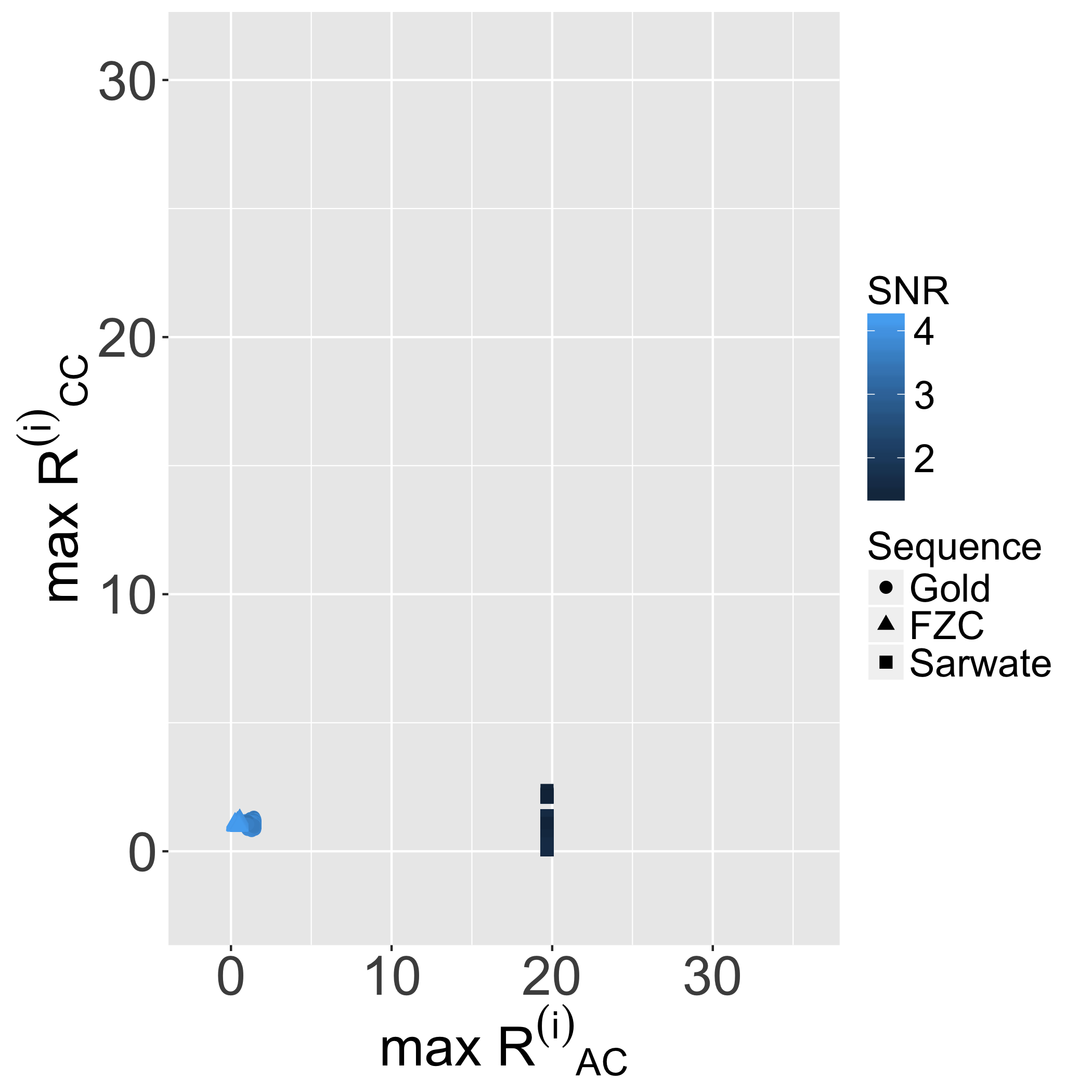} 
   \caption{Other sequences: maximum mean-square correlation in all the users: $K=4, Z_{AC}=2, Z_{CC}=1$}
   \label{fig:4_2_1_other_rmax}
\end{figure}

\newpage
\section{Figures of Numerical Result in Problem $(P2)$}

\begin{figure}[htbp] 
   \centering
   \includegraphics[width=2.3in]{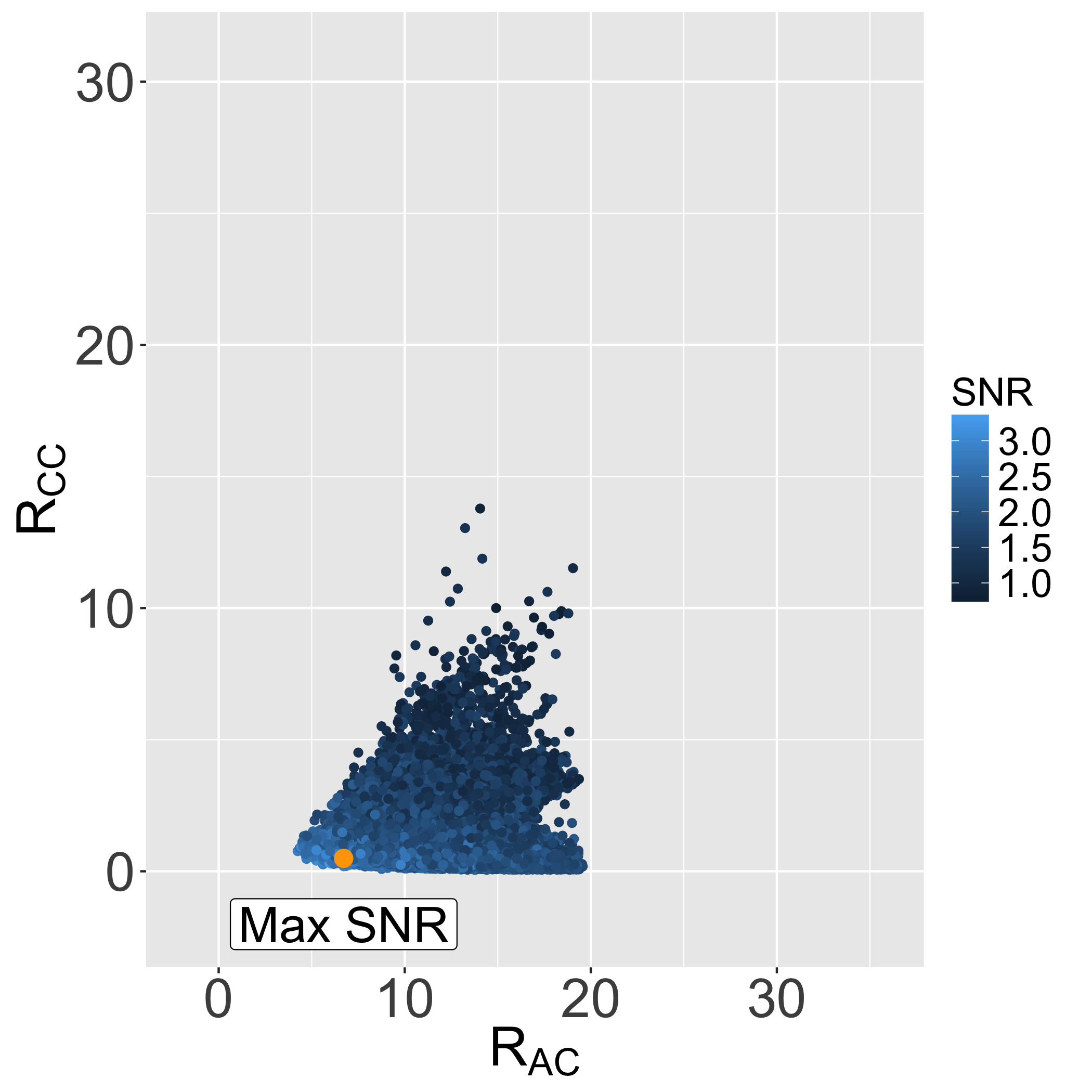} 
   \caption{Evaluation of solutions: mean-square correlation: $K=4, Z_{AC}=1, Z_{CC}=2$}
   \label{fig:4_1_2_r_min}
\end{figure}

\begin{figure}[htbp] 
   \centering
   \includegraphics[width=2.3in]{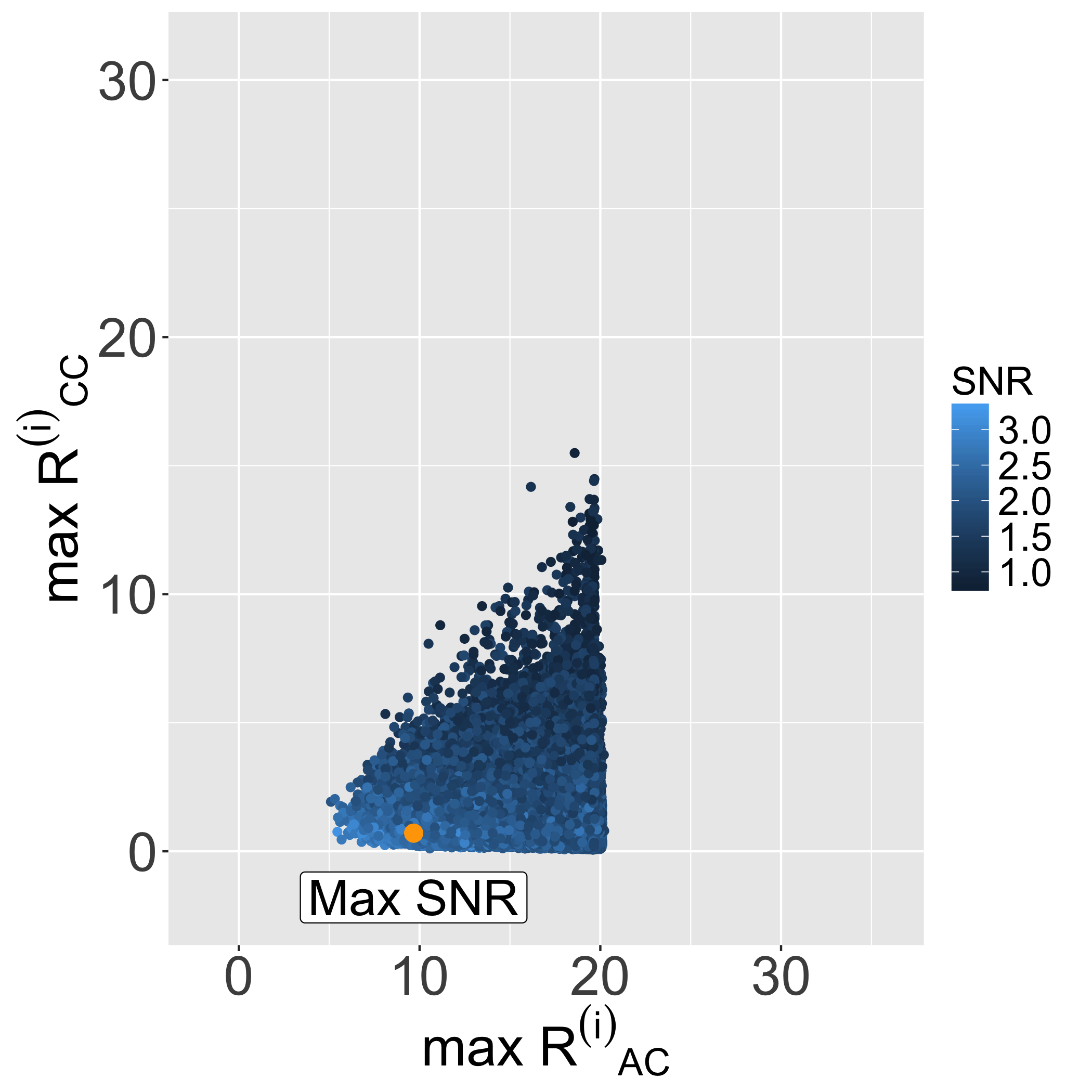} 
   \caption{Evaluation of solutions: maximum mean-square correlation in all the users: $K=4, Z_{AC}=1, Z_{CC}=2$}
   \label{fig:4_1_2_rmax_min}
\end{figure}

\begin{figure}[htbp] 
   \centering
   \includegraphics[width=2.3in]{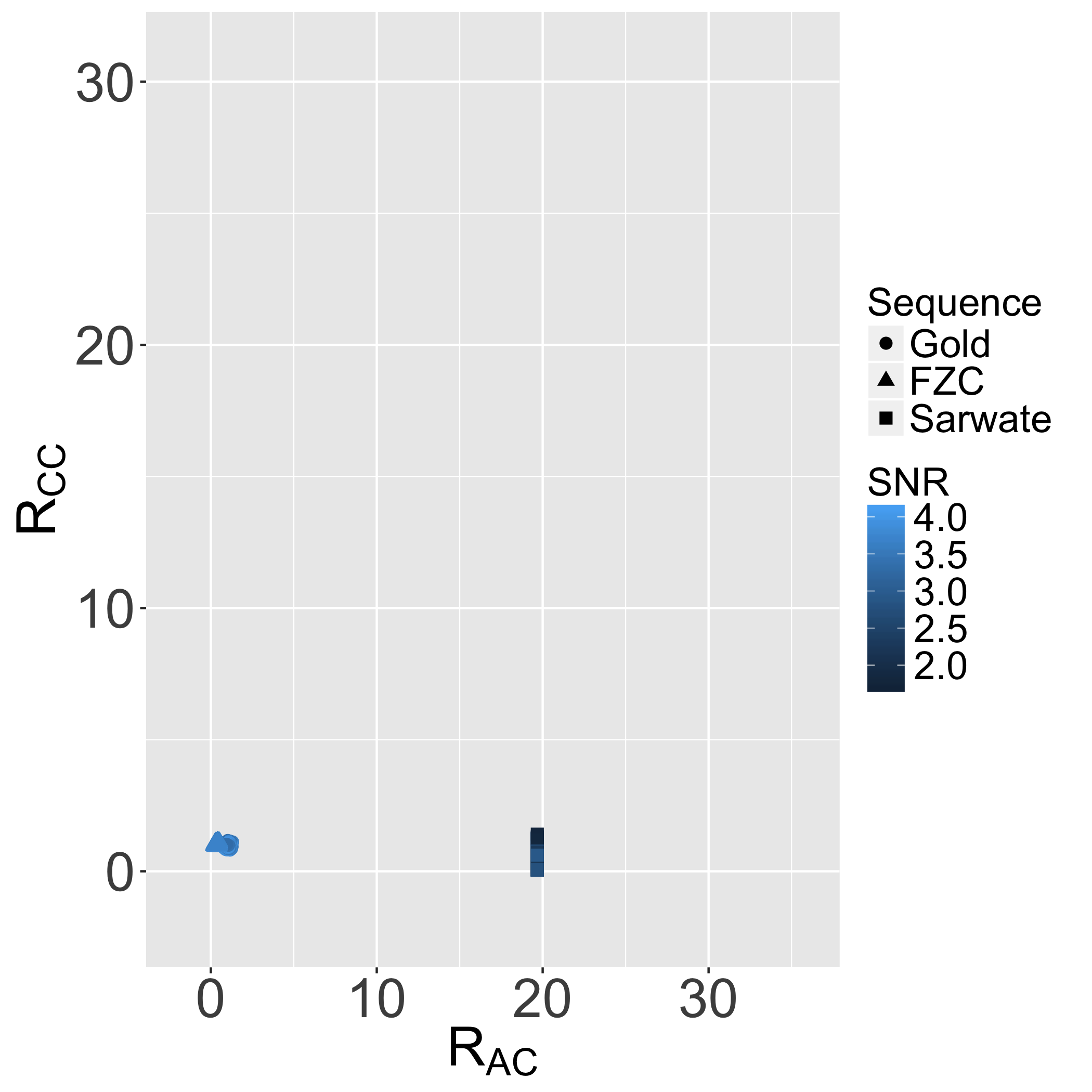} 
   \caption{Other sequences: mean-square correlation: $K=4, Z_{AC}=1, Z_{CC}=2$}
   \label{fig:4_1_2_other_r_min}
\end{figure}

\begin{figure}[htbp] 
   \centering
   \includegraphics[width=2.3in]{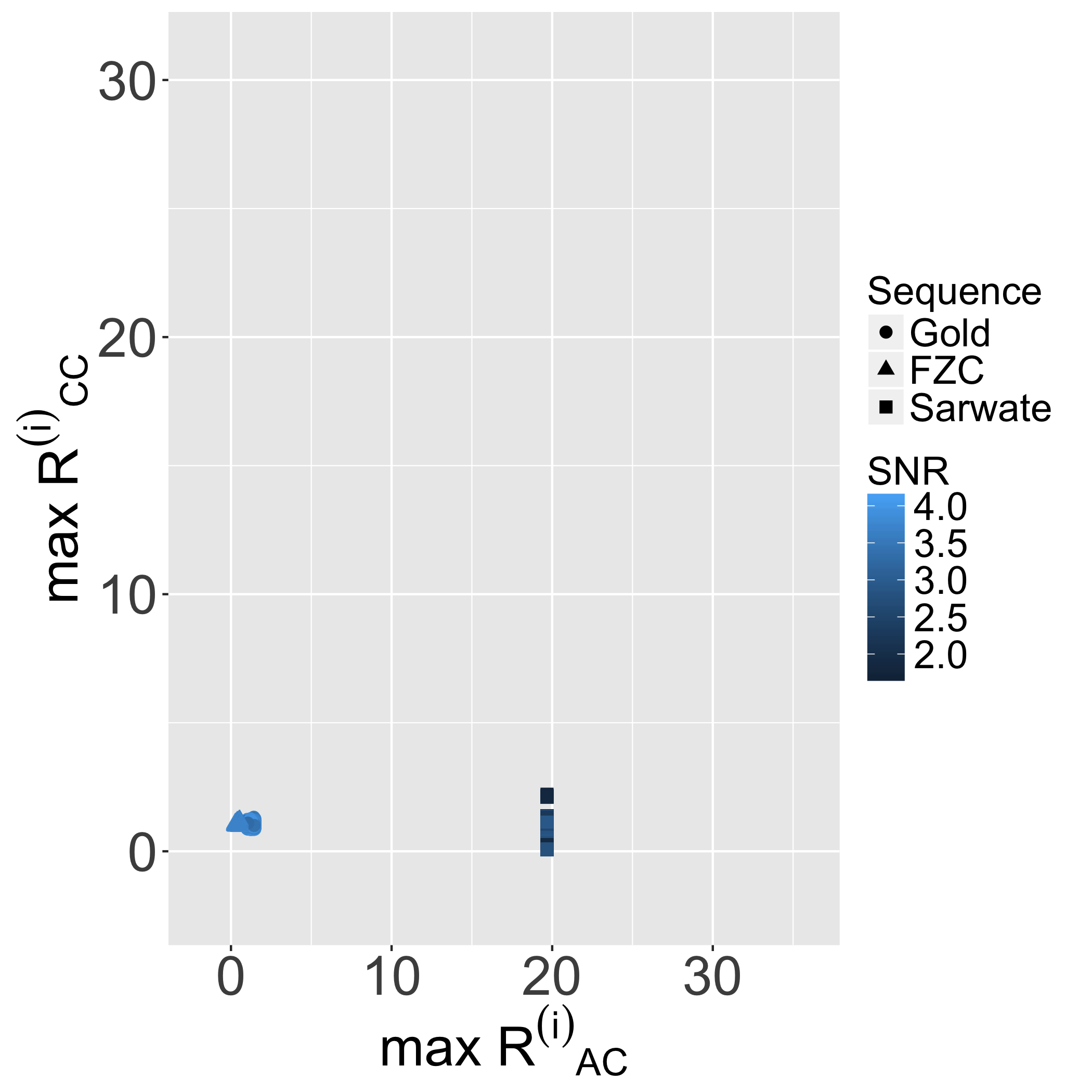} 
   \caption{Other sequences: maximum mean-square correlation in all the users: $K=4, Z_{AC}=1, Z_{CC}=2$}
   \label{fig:4_1_2_other_rmax_min}
\end{figure}

\begin{figure}[htbp] 
   \centering
   \includegraphics[width=2.3in]{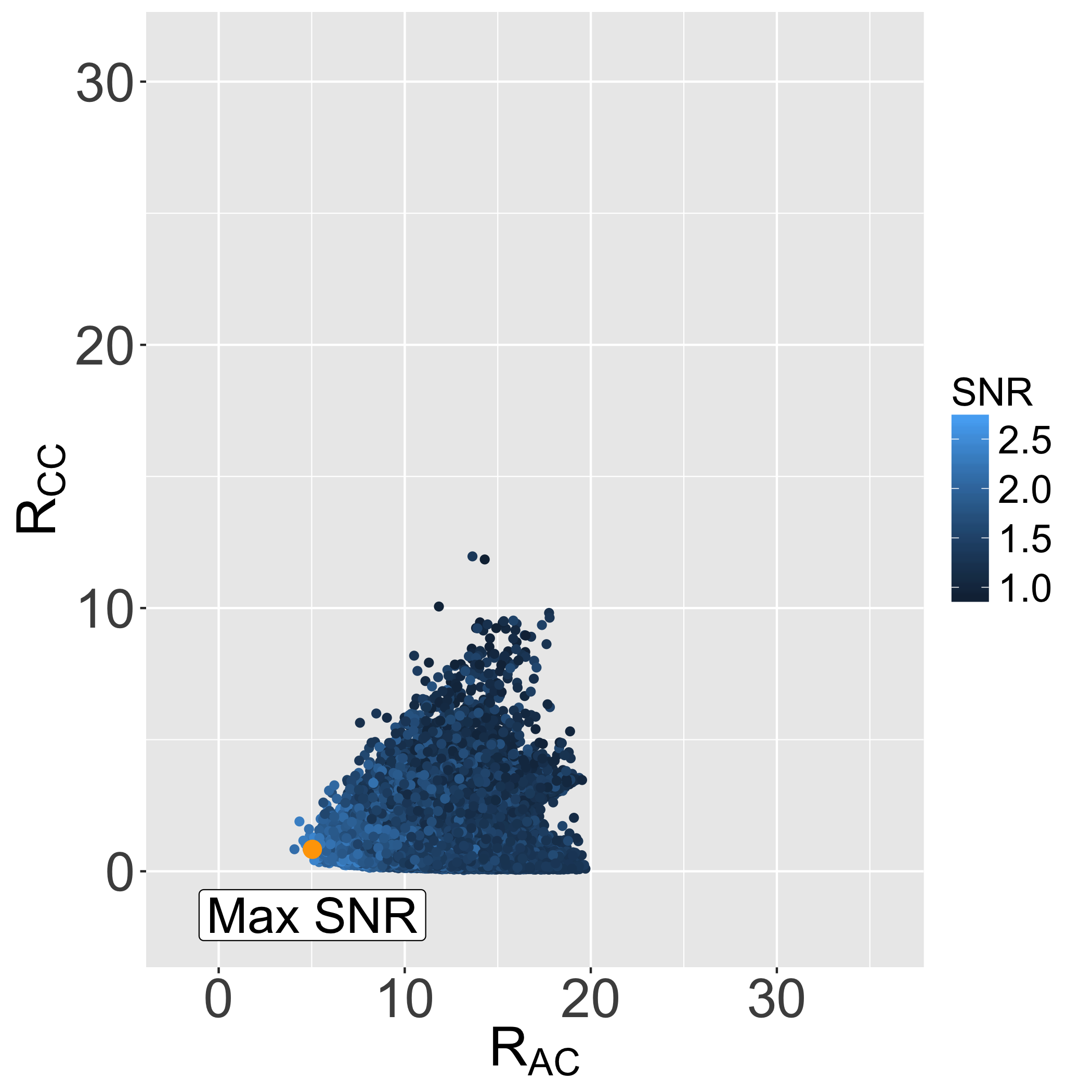} 
   \caption{Evaluation of solutions: mean-square correlation: $K=4, Z_{AC}=2, Z_{CC}=1$}
   \label{fig:4_2_1_r_min}
\end{figure}

\begin{figure}[htbp] 
   \centering
   \includegraphics[width=2.3in]{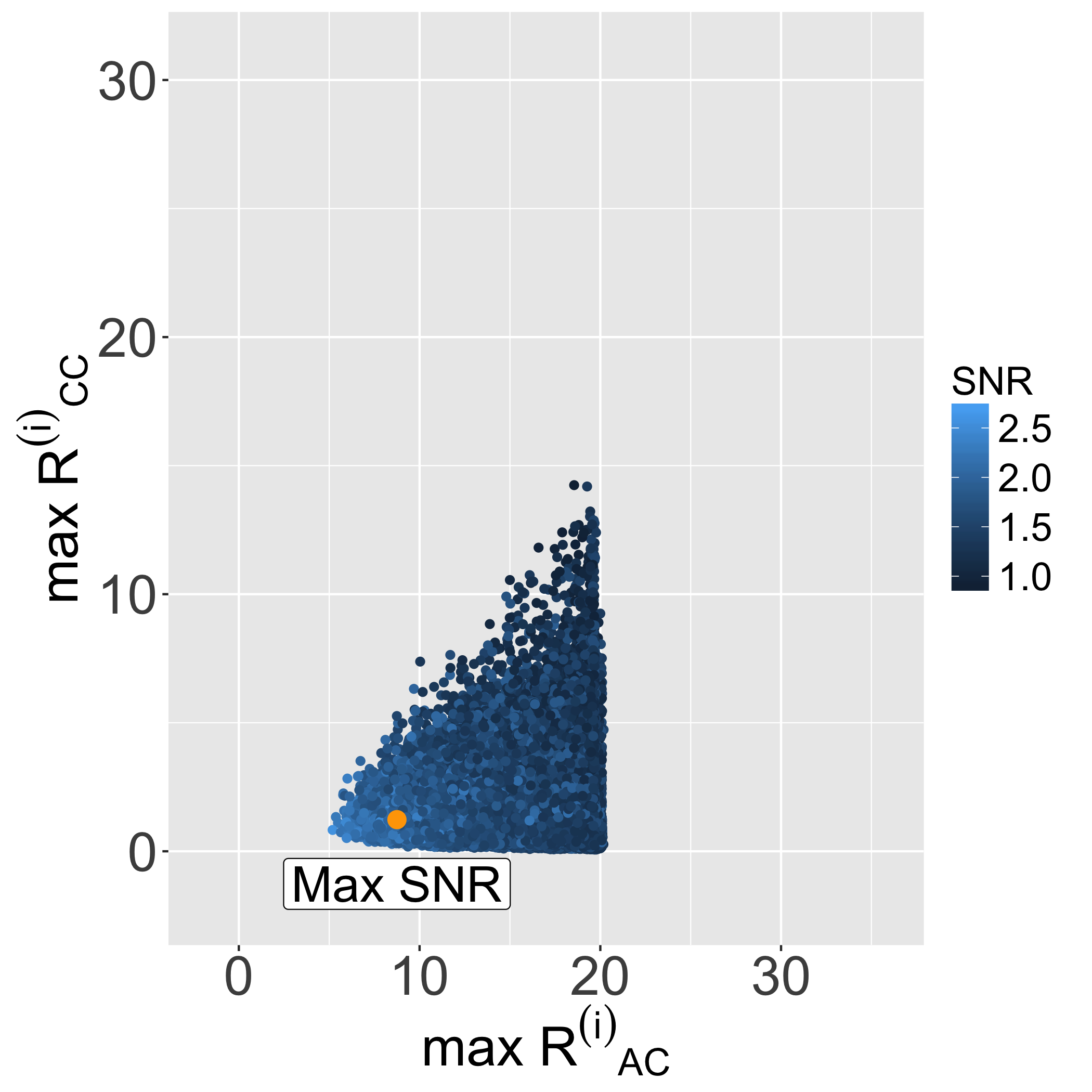} 
   \caption{Evaluation of solutions: maximum mean-square correlation in all the users: $K=4, Z_{AC}=2, Z_{CC}=1$}
   \label{fig:4_2_1_rmax_min}
\end{figure}

\begin{figure}[htbp] 
   \centering
   \includegraphics[width=2.3in]{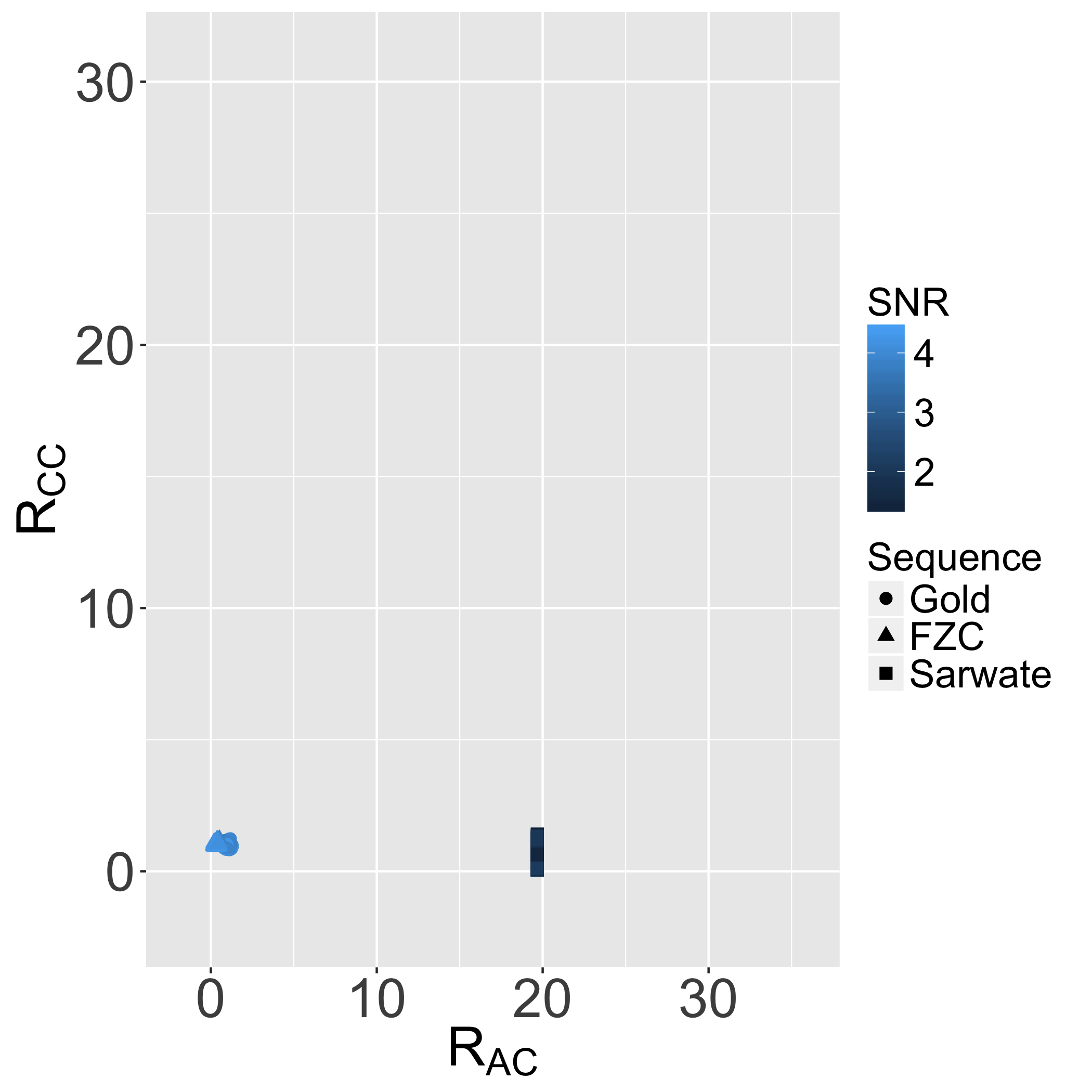} 
   \caption{Other sequences: mean-square correlation: $K=4, Z_{AC}=2, Z_{CC}=1$}
   \label{fig:4_1_1_other_r_min}
\end{figure}

\begin{figure}[htbp] 
   \centering
   \includegraphics[width=2.3in]{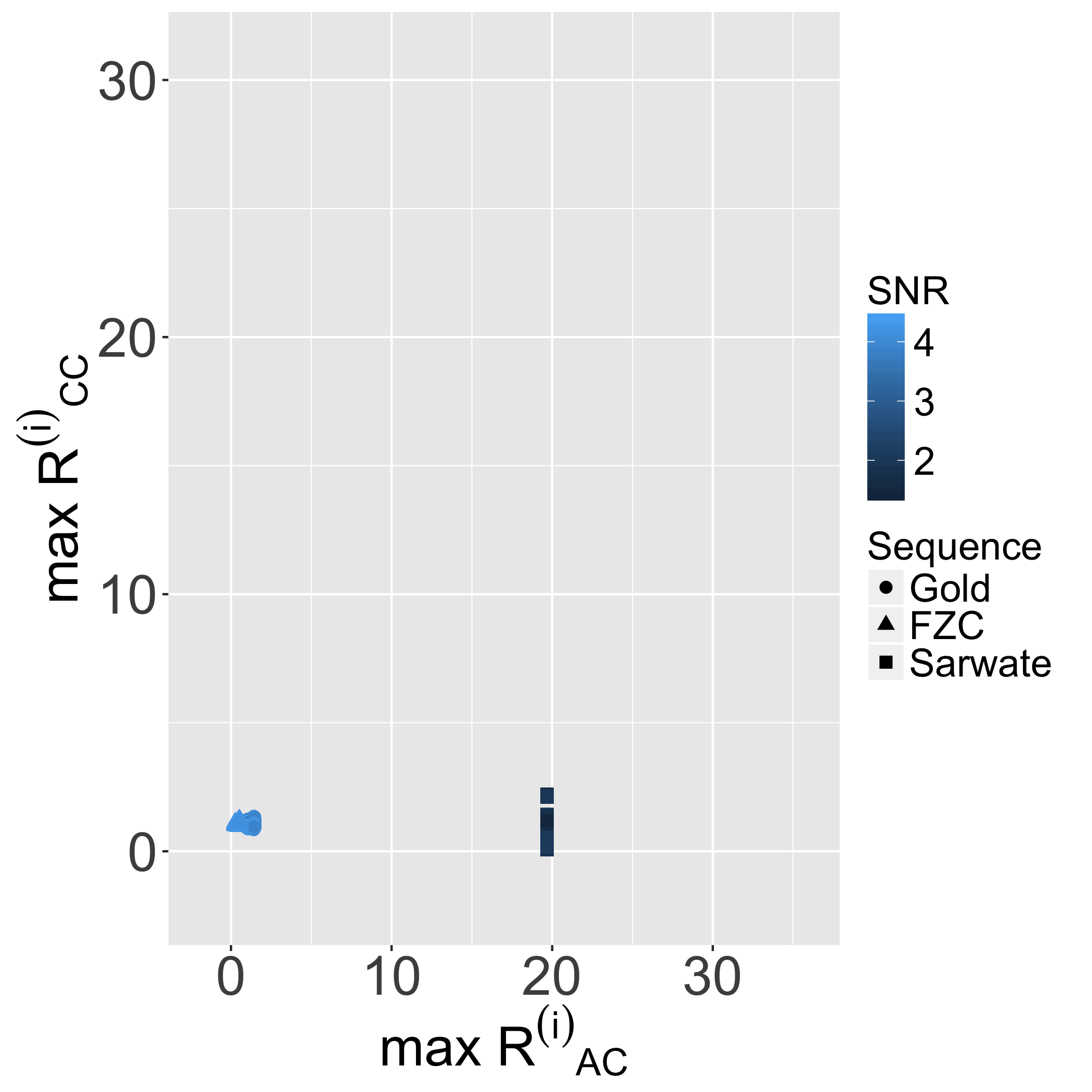} 
   \caption{Other sequences: maximum mean-square correlation in all the users: $K=4, Z_{AC}=2, Z_{CC}=1$}
   \label{fig:4_2_1_other_rmax_min}
\end{figure}

\newpage
\section*{Acknowledgment}
One of the authors, Hirofumi Tsuda, would like to thank for advice of Dr. Shin-itiro Goto.

\begin{thebibliography}{99}
\bibitem{dscdma} J. Proakis, ``Digital Communications. 1995'', McGraw-Hill, New York.
\bibitem{multiple} R. Steele and L. Hanzo, ``Mobile Radio Communications'', Second and Third Generation Cellular and WATM Systems: 2nd. IEEE Press-John Wiley, 1999.
\bibitem{shannon} C. E. Shannon, ''A mathematical theory of communication, ``Bell System Technical Journal, Volume 27, Issue 3, 379-423 (1948).
\bibitem{efficiency} S. Verd\'u and S. Shamai, ``Spectral efficiency of CDMA with random spreading.'' {\it IEEE Transactions on Information Theory} 45.2 (1999): 622-640.
\bibitem{pursley}M. B. Pursley, ``Performance evaluation for phase-coded spread-spectrum multiple-access communication. I-system analysis.'' {\it IEEE Transactions on Communications} 25 (1977): 795-799..
\bibitem{gold}R. Gold, ``Optimal binary sequences for spread spectrum multiplexing'', {\it IEEE Transactions on Information Theory},
 13.4 (1967): 619-621.
 \bibitem{chaos_cdma} G. Heidari-Bateni and C. D. McGillem. ``A chaotic direct-sequence spread-spectrum communication system.'' {\it IEEE Transactions on communications} 42.234 (1994): 1524-1527.
\bibitem{chaos_mod} K.S. Halle, C.W. Wu, M. Itoh and L.O. Chua. ``Spread spectrum communication through modulation of chaos.'' {\it International Journal of Bifurcation and Chaos} 3.02 (1993): 469-477.
\bibitem{logistic} Y. Soobul, K. Chady and H. C.S. Rughooputh. ``Digital chaotic coding and modulation in CDMA.'' {\it Africon Conference in Africa, 2002. IEEE AFRICON. 6th}. Vol. 2. IEEE, 2002.
\bibitem{ergotic} C.C. Chen, K. Yao, K. Umeno and E. Biglieri, ``Design of spread-spectrum sequences using chaotic dynamical systems and ergodic theory.'' {\it IEEE Transactions on Circuits and Systems I: Fundamental Theory and Applications} 48.9 (2001): 1110-1114.
\bibitem{autocor} G. Mazzini, R. Rovatti and G. Setti. ``Interference minimisation by autocorrelation shaping in asynchronous DS-CDMA systems: chaos-based spreading is nearly optimal.'' {\it Electronics Letters} 35.13 (1999): 1054-1055.
\bibitem{mazzini} G. Mazzini, G. Setti and R. Rovatti. ``Chaotic complex spreading sequences for asynchronous DS-CDMA. I. System modeling and results.''  {\it IEEE Transactions on Circuits and Systems I: Fundamental Theory and Applications}, 44.10 (1997): 937-947.
\bibitem{kaddoum} G. Kaddoum, M. Coulon, D. Roviras and P. Charg\'e. ``Theoretical performance for asynchronous multi-user chaos-based communication systems on fading channels.'' {\it Signal Processing} 90.11 (2010): 2923-2933.
\bibitem{multipath} G. Kaddoum, D Roviras, P Charg\'e and D. Fournier-Prunaret. ``Accurate bit error rate calculation for asynchronous chaos-based DS-CDMA over multipath channel.'' {\it EURASIP Journal on Advances in Signal Processing} 2009 (2009): 48.
\bibitem{indoor} R. Takahashi and K. Umeno. ``Performance evaluation of CDMA using chaotic spreading sequence with constant power in indoor power line fading channels.'' {\it IEICE Transactions on Fundamentals of Electronics}, Communications and Computer Sciences 97.7 (2014): 1619-1622.
\bibitem{nonlinear} D. P. Bertsekas, ``Nonlinear programming'', Belmont: Athena scientific, 1999.
\bibitem{KKT}W. Kuhn and A. W. Tucker, ``Nonlinear programming”, in J. Neyman (ed.), {\it Proceedings of the Second Berkley Symposium on Mathematical Statistics and Probability} (University of California Press, Berkley, CA), pp. 481-492, 1951.
\bibitem{sqp} P. T. Boggs and J. W. Tolle, ``Sequential quadratic programming'', {\it Acta Numerica} 4 (1995): 1-51.
\bibitem{nop} A. W\"{a}chter and L. T. Biegler. ``On the implementation of an interior-point filter line-search algorithm for large-scale nonlinear programming.'' {\it Mathematical Programming} 106.1 (2006): 25-57.
\bibitem{bfgs} D. C. Liu, and J. Nocedal. ``On the limited memory BFGS method for large scale optimization.'' {\it Mathematical programming} 45.1-3 (1989): 503-528.
\bibitem{ncg} S. G. Nash, "A survey of truncated-Newton methods.'' {\it Journal of Computational and Applied Mathematics} 124.1 (2000): 45-59.
\bibitem{slsqp} D. Kraft, ``A software package for sequential quadratic programming''. Obersfaffeuhofen, Germany: DFVLR, 1988.
\bibitem{part_I} H. Tsuda, and K. Umeno, ``Non-Linear Programming: Maximize SNR for Designing Spreading Sequence -- Part I: SNR versus Mean-Square Correlation'', arXiv:1612.08232 (2016)
\bibitem{fzcfam} J. Oppermann and S.V. Branka. ``Complex spreading sequences with a wide range of correlation properties." {\it IEEE Transactions on Communications} 45.3 (1997): 365-375.
\bibitem{meansquare} K. H. A. K\"arkk\"ainen,  ``Mean-square cross-correlation as a performance measure for department of spreading code families." {\it Spread Spectrum Techniques and Applications, 1992. ISSTA 92. IEEE Second International Symposium on}. IEEE, 1992.
\bibitem{basis} H. Tsuda and K. Umeno, ``Orthogonal Basis Spreading Sequence for Optimal CDMA'', {\it JSIAM Letters}, Vol. 8  (2016): 77-80. 
\bibitem{cmgc} E. Telatar. ``Capacity of Multi‐antenna Gaussian Channels." {\it European transactions on Telecommunications} 10.6 (1999): 585-595.
\bibitem{minimax} A. Vardi, ``New minimax algorithm." {\it Journal of Optimization Theory and Applications} 75.3 (1992): 613-634.
\bibitem{zadoff} R. Frank, S. Zadoff and R. Heimiller. ``Phase shift pulse codes with good periodic correlation properties (corresp.)." {\it IRE Transactions on Information Theory} 8.6 (1962): 381-382.
\bibitem{chu}D. Chu. ``Polyphase codes with good periodic correlation properties (corresp.)." {\it IEEE Transactions on information theory} 18.4 (1972): 531-532.
\bibitem{sarwate} D. V. Sarwate, ``Bounds on crosscorrelation and autocorrelation of sequences", {\it IEEE Transactions on Information Theory}, 25.6 (1979): 720-724.
\bibitem{barrier} C. W. Carroll, ``The created response surface technique for optimizing nonlinear, restrained systems." {\it Operations Research} 9.2 (1961): 169-184.
\bibitem{penalty} \"O. Yeniay, ``Penalty function methods for constrained optimization with genetic algorithms." {\it Mathematical and Computational Applications} 10.1 (2005): 45-56.
\end{thebibliography}
\end{document}